\documentclass[12pt,preprint]{aastex}

\newcommand{\lam}{$\lambda$}

\newcommand{\ecsa}{erg cm$^{-2}$ s$^{-1}$ sr$^{-1}$ \AA$^{-1}$} 
\newcommand{\kms}{km~s$^{-1}$}
\newcommand{\as}{${^\prime}{^\prime}$}
\newcommand{\tef}{{$T_{eff}$ }}

\newcommand{\etal}{{\em et al.\ }}

\newcommand{\ionx}[2]{#1\,{\sc #2}}

\def\ion[#1 #2]{#1\,{\sc #2}}
\def\ergs[#1]{#1 {ergs}~{cm$^{-2}$}\,{s$^{-1}$}\,{sr$^{-1}$}}
\def\dens[#1]{10$^{#1}$\hskip 1.5pt{cm$^{-3}$}}
\def\densr[#1 #2]{10$^{#1}$\hskip 1pt{--}\hskip .5pt{10$^{#2}$}\hskip 1.5pt{cm$^{-3}$}}
\def\fl[#1 #2]{{#1}$\pm${#2}}
\def\orb[#1 #2]{{$#1^{#2}$}}
\def\ls[#1 #2]{{$^{#1}${#2}}}
\def\tm[#1 #2 #3]{{$^{#1}${#2}$_{#3}$}}

\begin{document}

\title{CHIANTI -- An atomic database for emission lines. X. Spectral atlas of a cold feature observed with Hinode/EIS}

\author{E. Landi\altaffilmark{1} \and P. R. Young,\altaffilmark{1,2}}

\altaffiltext{1}{Naval Research Laboratory, Space Science Division, Washington, DC 20375}
\altaffiltext{2}{George Mason University, 4400 University Drive, Fairfax, VA 22030}

\begin{abstract}
In this work we report on a cold, bright portion of an active region observed by 
EIS. The emitting plasma was very bright at transition region
temperatures, and the intensities of lines of ions formed between $10^5-10^6$~K were 
enhanced over normal values. The data set constitutes an excellent laboratory where
the emission of transition region ions can be tested. We first determine the thermal 
structure of the observed plasma, and then we use it 1) to develop a spectral atlas,
and 2) to assess the quality of CHIANTI atomic data by comparing 
predicted emissivities with observed intensities. We identify several lines never
observed before in solar spectra, and find an overall excellent agreement between
CHIANTI predicted emissivities and observations.
\end{abstract}


\section{Introduction}

The CHIANTI atomic database \citep{dere97,dere09} provides
up-to-date, assessed atomic data 
for most astrophysically useful ions as well as software for deriving
emission line emissivities and synthetic spectra. It has been used to
model and interpret emission from a wide range of objects in
astrophysics including the Sun's 
outer atmosphere, the Jupiter-Io plasma torus \citep{steffl08}, T
Tauri stars \citep{gunther08}, the interstellar medium
\citep{sallmen08} and supernova remnants \citep{reyes08}. A vital part
of maintaining CHIANTI is the assessment of data quality through
comparisons of the atomic models with observed spectra. The Sun's atmosphere
is a natural target for such studies as there have been several high
resolution spectrometers flown on both rockets and satellites that
have produced high signal-to-noise data over large wavelength ranges
in the ultraviolet and X-ray regions. In addition, the wide range of
structures offered by the Sun -- coronal holes, quiet Sun, active
regions, flares -- yield very different spectra that allow particular
atomic models to be studied in different physical conditions. Three
previous data assessments have been performed: the comparison of the
SERTS-89 rocket flight spectrum with version~1 of CHIANTI by
\citet{young98}; the \citet{landi02,landi02b} comparisons of version~3 of
CHIANTI with off-limb quiet Sun spectra obtained with the SUMER and CDS
instruments, respectively, on board the SOHO satellite; and the study of 
an X-ray spectrum obtained with the Flat Crystal Spectrometer on board 
the Solar Maximum Mission by \citet{landi06b} using version~5 of CHIANTI.

For the present work, an unusual spectrum obtained with the EUV Imaging
Spectrometer \citep[EIS,][]{culhane07} on board the Hinode satellite
\citep{kosugi07} has been found that shows strongly enhanced lines 
from the upper transition region corresponding to temperatures 
$\log\,(T/K)=5.0$--5.9. EIS takes high resolution spectra in the
wavelength ranges 170--211 and 246--291~\AA\ and the data represent an
excellent opportunity to study atomic physics properties of a group of
ions that normally emit weak lines, yet yield valuable information about 
the emitting plasma.

The paper is structured as follows. First, details of the observation and 
the procedure for extracting, calibrating, and fitting the spectrum are 
described. Using lines from all the ions observed by EIS, we determine a 
first, approximate, differential emission measure (DEM) curve. This is 
used, together with the L-function method of Landi \& Landini (1997), to 
identify blends or atomic physics issues and select a set of lines free 
from problems. These are used to determine a more accurate DEM curve. 
This new curve is used 1) to derive a synthetic spectrum that is used 
to confirm line identifications in the atlas, and 2) to compare CHIANTI 
emissivities and observed line intensities for all the ions in the 
$\log\,T=5.0$--5.9 temperature range that have lines identified in the 
atlas.

The comparison between CHIANTI emissivities and EIS observations will be
split between three separate papers. In the present paper we will consider
all elements except iron. In a second paper \citep{young09a} we will
discuss the three iron ions \ionx{Fe}{vii--ix}
whose emission is very prominent in the EIS observations we use here, but
require special attention due to the large number of lines and of new
identifications we made. In a third paper we will carry out the comparison
between the CHIANTI emissivities for coronal ions and another set of
observations, carried out with a special observing sequence and on a solar
target specifically chosen to enhance coronal emission. Thus, we will not
consider coronal ions in the present dataset.

\section{Observation}

\begin{figure}
\includegraphics[width=11.0cm,height=17.0cm,angle=90]{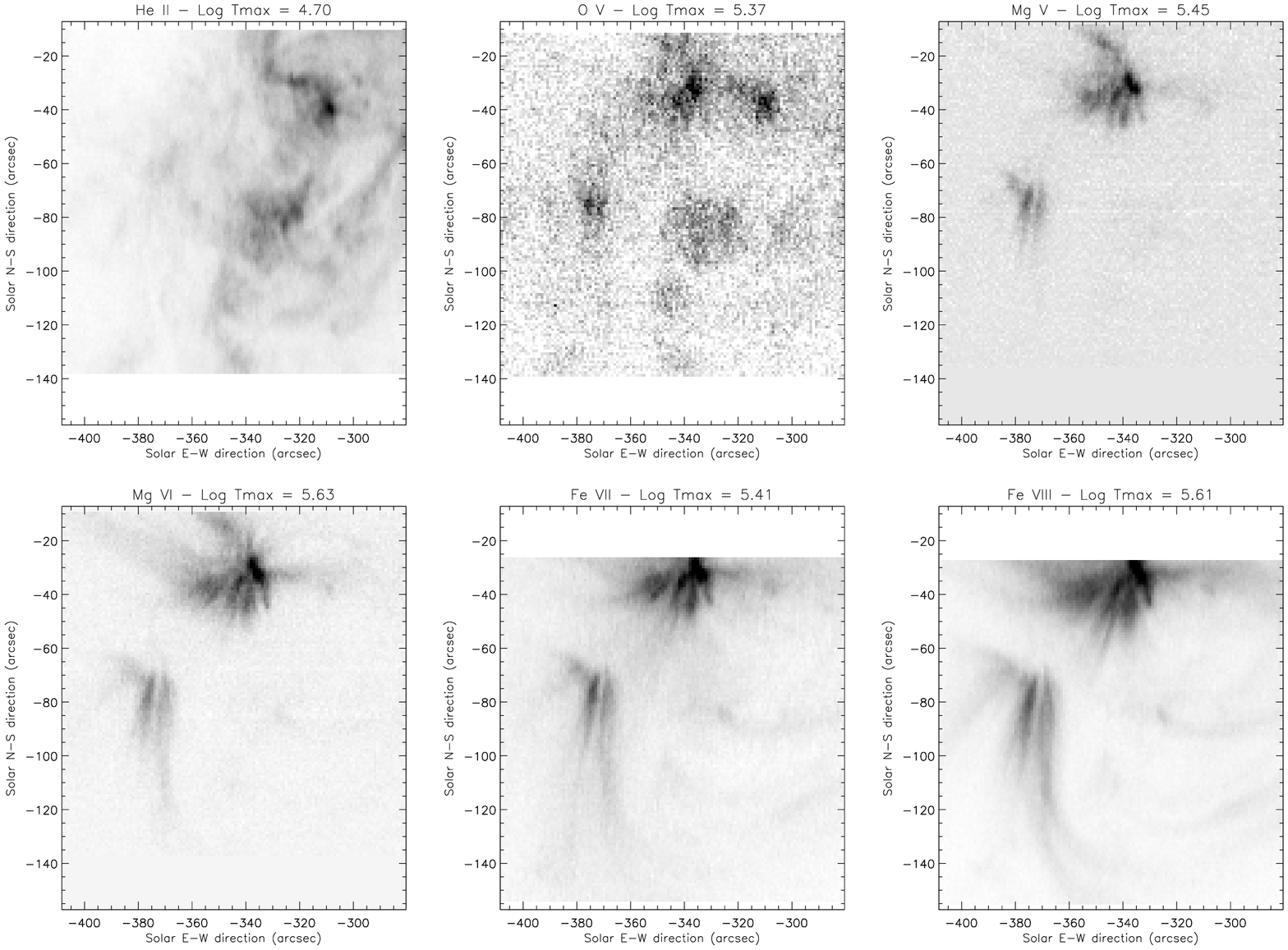}
\includegraphics[width=11.0cm,height=17.0cm,angle=90]{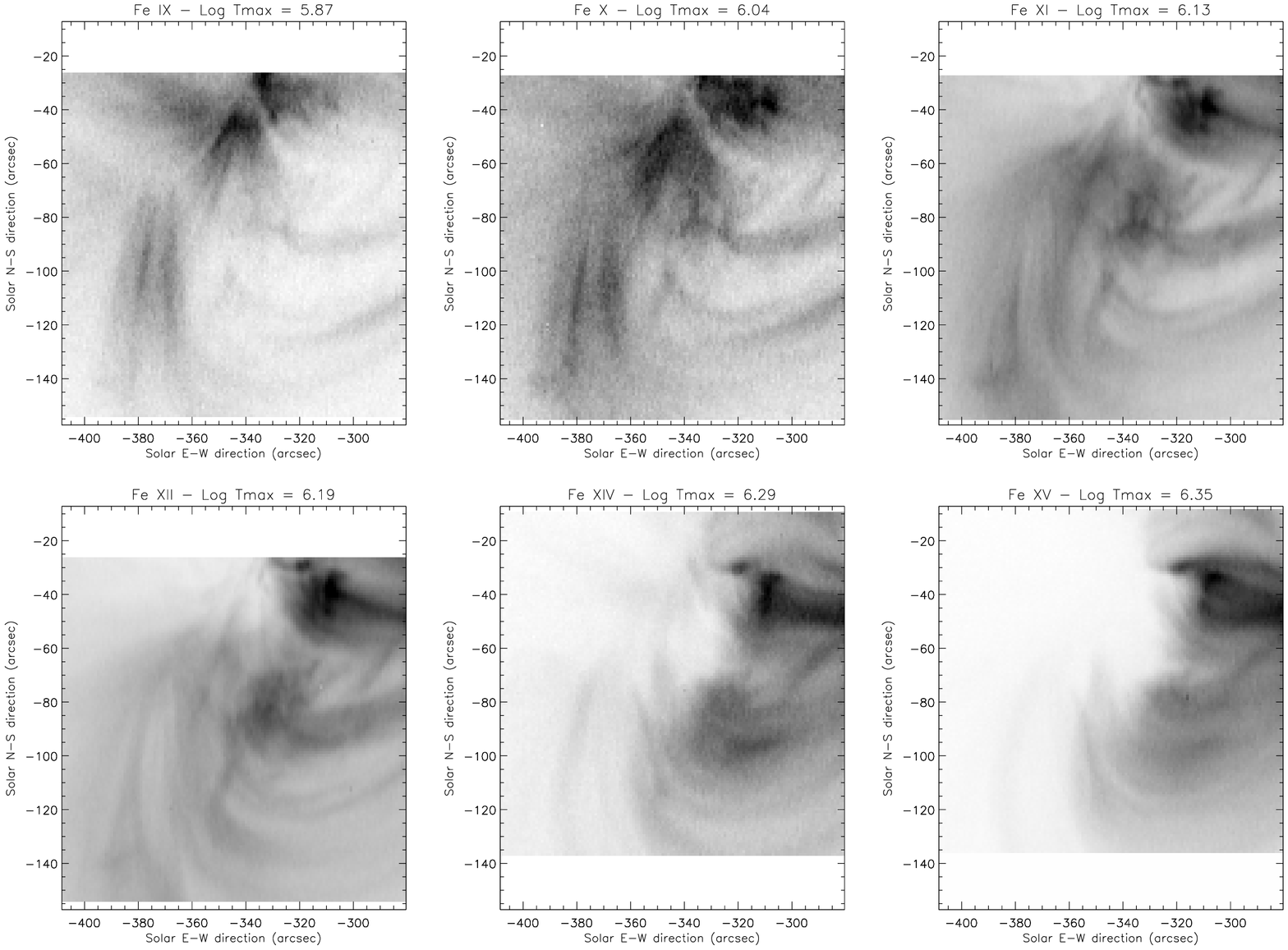}
\caption{\label{maps} Intensity maps for bright lines formed at different temperature
regimes. The temperature of maximum abundance, from the CHIANTI~6 ionization equilibrium
(Dere \etal 2009) is reported as $\log T_{max}$ for each line.}
\end{figure}

The dataset studied here is the same as that analysed by
\citet{young09} -- a single EIS raster obtained on 2007 February 21
and pointed at active region AR 10942. The complete EIS spectral range
is obtained over a 128\as\ $\times$ 128\as\ spatial area with a 25~s
exposure at each slit position. \citet{young09} selected a spatial
area where newly-identified lines of \ion[Fe ix] were espectially
strong. For the present work we choose a region of 30 pixels that
corresponds to a bright point apparent in \ion[Fe viii] images 
centered at X=-335\arcsec, Y=-30\arcsec (Fig.~\ref{maps}). The 
bright point appears to be related to
the footpoint regions of coronal loops. A SOT magnetogram obtained at
02:00~UT shows that the bright point lies within a unipolar plage
region. 

The data were calibrated using the standard calibration routine
EIS\_PREP which is available in the Solarsoft software
distribution. EIS\_PREP is described in detail by \citet{young08}, and
the routine has been expanded since that work with the following
features. Anomalously bright pixels referred to as warm pixels are now
directly removed by EIS\_PREP through comparison with warm pixel maps
obtained by regular engineering studies. For the present observation
the warm pixel map was obtained on 2007 March 3. Previously, warm
pixels were flagged via the cosmic ray detection algorithm. For full
CCD spectra such as those analysed here the method for estimating the
CCD background (consisting of the pedestal and dark current) is
different to that described by \citet{young08}. The two
2048$\times$1024 CCDs that
measure the two different EIS wavelength bands are each read out as
two halfs of size 1024$\times$1024, the four `halfs' being referred to
as quadrants. For each quadrant an area of 46 pixels wide in the
CCD X-direction (corresponding to wavelength) has been identified as
being relatively free of emission lines. In each case these areas are
where the effective area of the instrument is low. The median value of
the data number (DN) values of each pixel in these 46 pixel wide
regions are treated as the CCD background for that quadrant, and thus
subtracted from the data by EIS\_PREP.

As the intrinsic EUV spectrum background is very low in the EIS
wavelength bands, a consequence of this method of background
subtraction for full CCD data is that a large number of pixels (up to 50~\%) can end up
with a zero or negative DN value. By default EIS\_PREP treats such
data points as `missing' data since it is not possible to assign a photon
statistics error to them. However, by specifying the keyword /RETAIN
the software will assign an error to these points that is simply the
estimated dark current error estimate and treat the photon statistics error
as zero. This is the option that has been used for the present work. 

Following the calibration by EIS\_PREP, the next step is to average
the spectra from the 30 pixel spatial region of interest to yield a
single spectrum for analysis. This procedure is complicated by both
the CCD spatial offset \citep{young07a} and the
EIS spectrum tilt identified by \citet{young08}, whereby a given
spatial feature appears at different CCD Y-positions depending on the
wavelength. The routine EIS\_CCD\_OFFSET in the EIS software tree
yields the spatial offset relative to \ion[He ii] \lam256.32 as a
function of wavelength. This routine uses the spectrum tilt gradient
derived by \citet{young08} for the EIS short wavelength (SW) band and
assumes that the same tilt applies to the long wavelength (LW)
band. In addition the offset between the SW and LW bands was derived
by co-aligning images in \ion[Fe viii] \lam185.21 and \ion[Si vii]
\lam275.35, which are observed to be very similar to each other. The
spatial offsets can be as large as 21 pixels for the shortest and
longest wavelength lines observed by EIS.

By specifying a spatial mask in the \ion[Fe viii] \lam194.66 line,
the routine EIS\_MASK\_SPECTRUM takes the level-1 FITS file output by
EIS\_PREP together with this mask and derives a single spectrum
averaged over the specified spatial region. The routine goes through
each wavelength in the spectrum and computes an adjusted pixel mask
based on the offset relative to \lam194.66 specified by
EIS\_CCD\_OFFSET. The pixels identified by this adjusted pixel mask
are then averaged to yield an intensity value for that wavelength.

\begin{figure}[h]
\epsscale{1.0}
\plotone{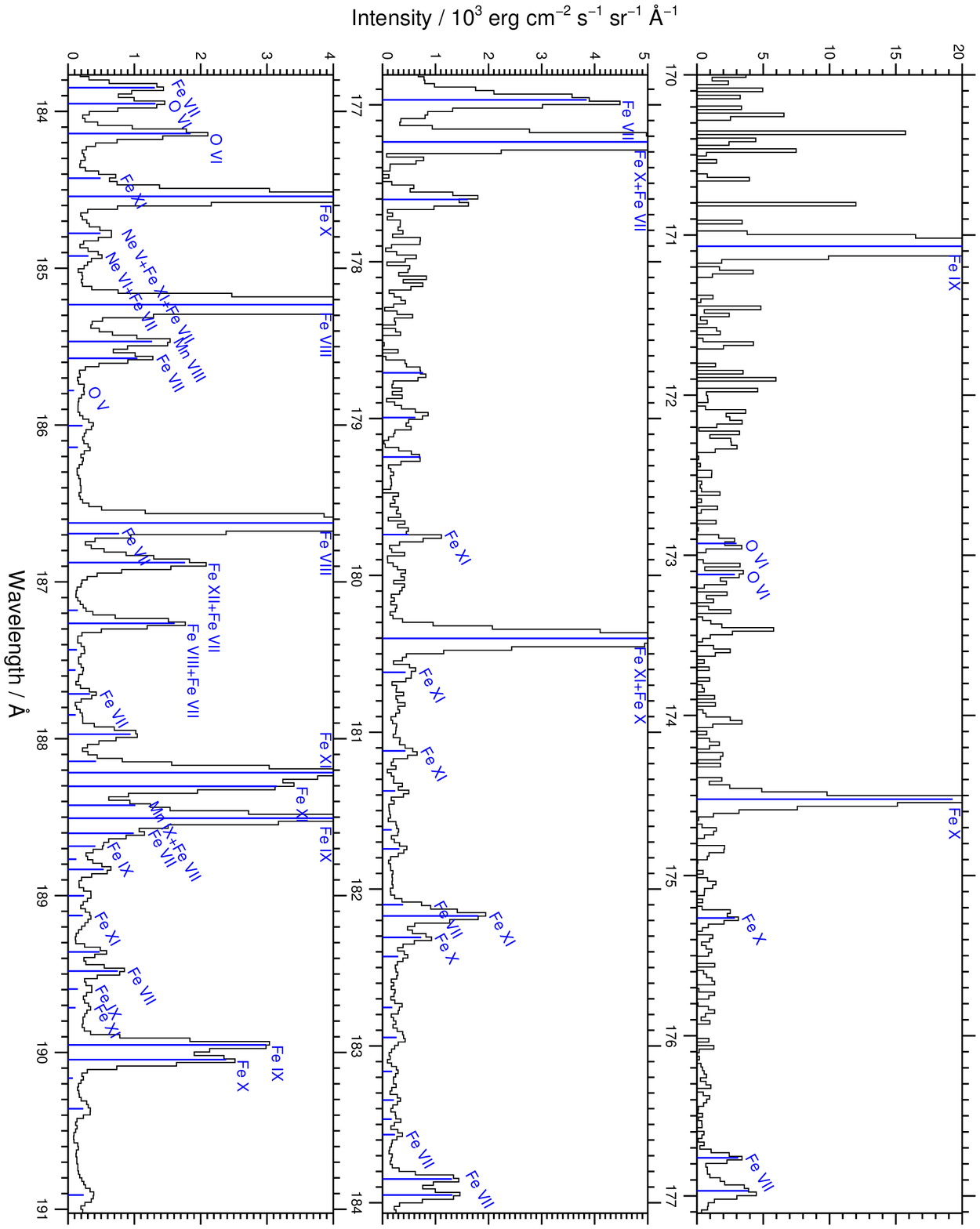}
\caption{Panels showing three sections of the EIS SW spectrum analyzed in
the present work. The spectrum is displayed in a black histogram plot,
and each line present in Table~\ref{linelist} is represented by a
vertical blue line, then length of which corresponds to the peak of
the fitted Gaussian. For identified lines, the emitting ion is
shown. The Y-scale varies for each panel according to the strengths of
the lines in the spectrum section. The spectral ranges of each panel
overlap with the neighbouring panels by around 0.3~\AA.}
\label{fig.spec1}
\end{figure}

\begin{figure}[h]
\epsscale{1.0}
\plotone{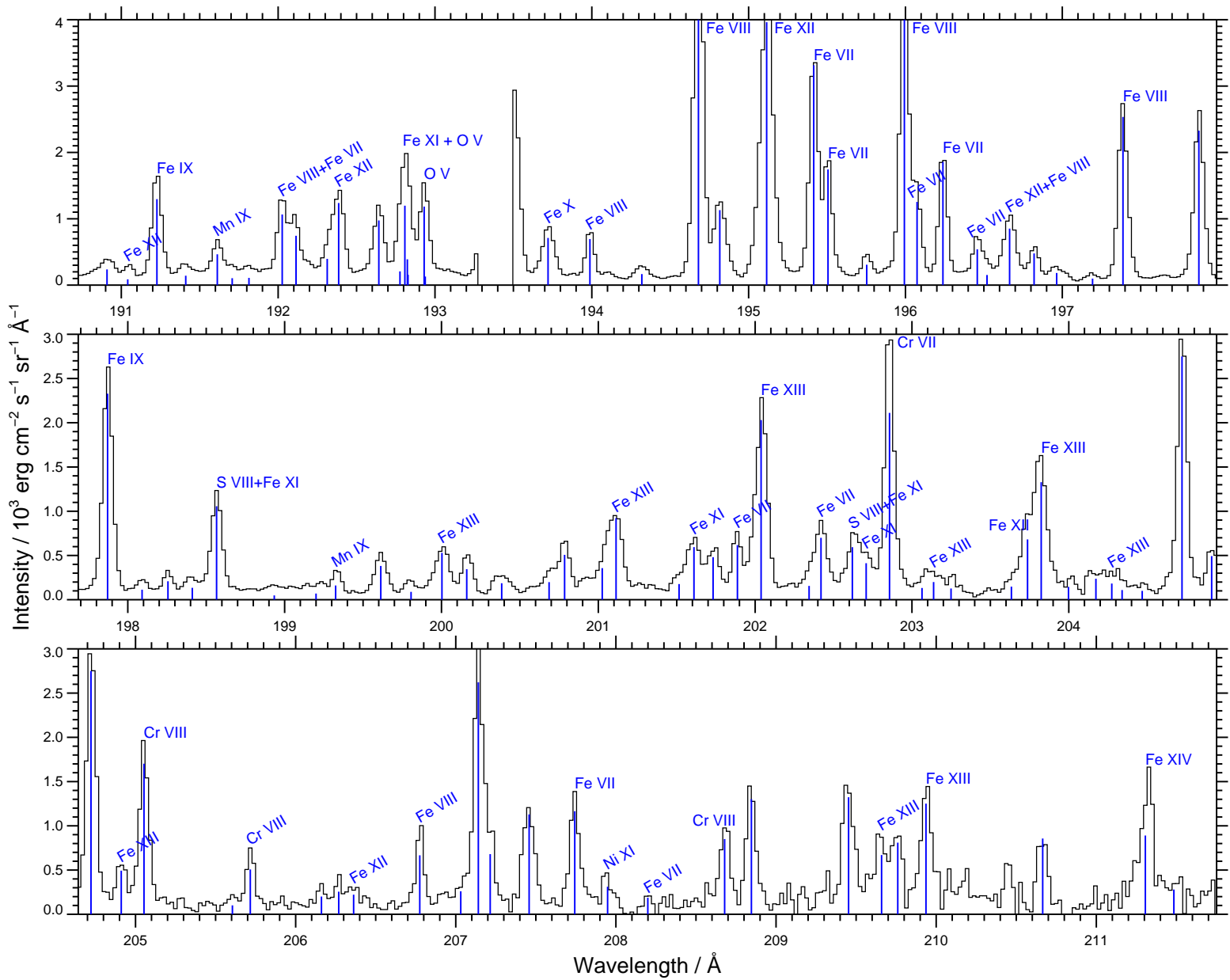}
\caption{Panels showing the second half of the EIS SW spectrum. See
  the caption of Fig.~\ref{fig.spec1} for more details. Note the gap
  around 193.4~\AA\ which is due to dust on the EIS detector blocking
  the spectral signal. The long wavelength half of the \ionx{Fe}{xii}
  \lam193.51 line profile can be seen at 193.5~\AA.}
\label{fig.spec2}
\end{figure}

\begin{figure}[h]
\epsscale{1.0}
\plotone{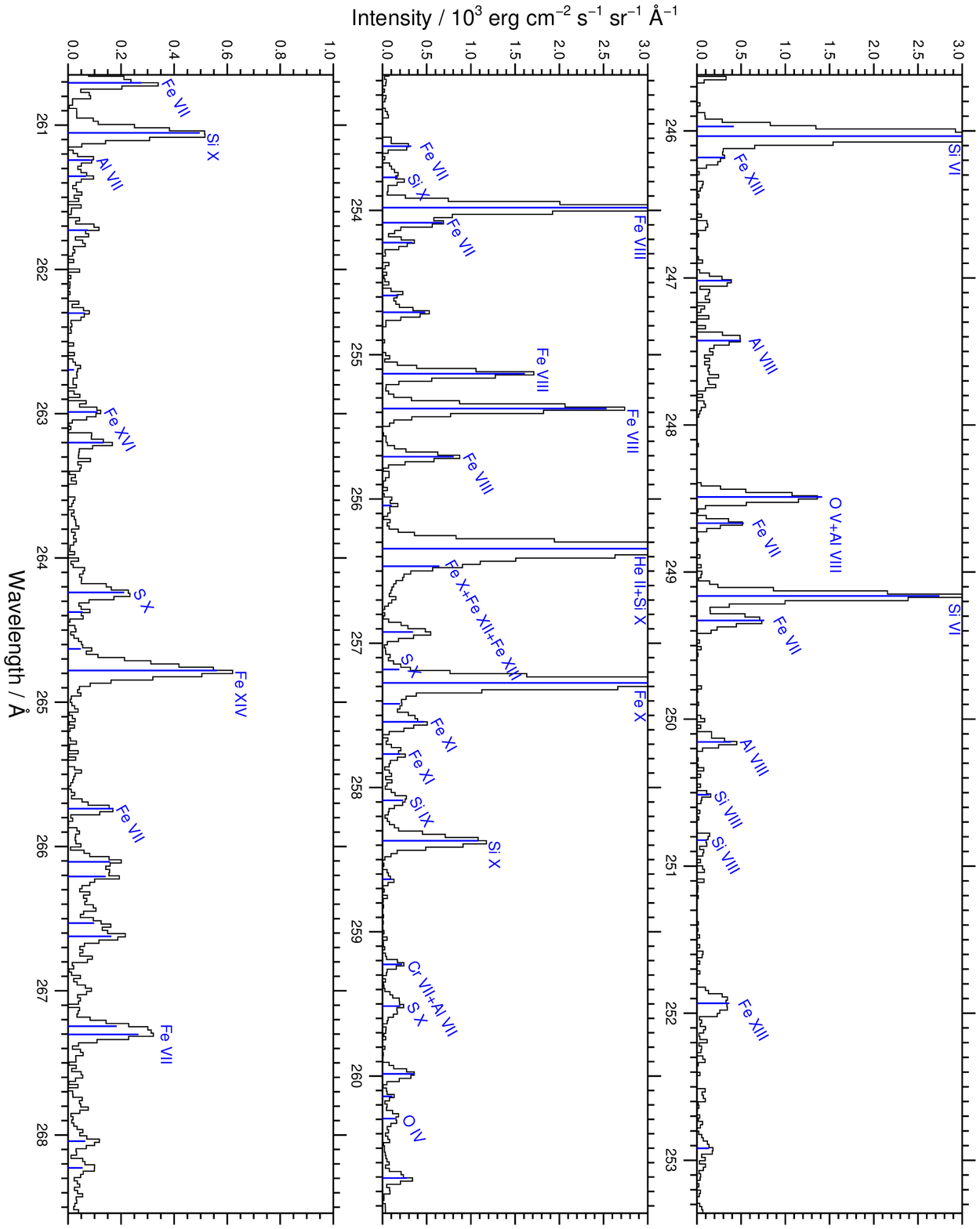}
\caption{Panels showing the short wavelength half of the EIS LW
  spectrum. See the caption of Fig.~\ref{fig.spec1} for more details.}
\label{fig.spec3}
\end{figure}

\begin{figure}[h]
\epsscale{1.0}
\plotone{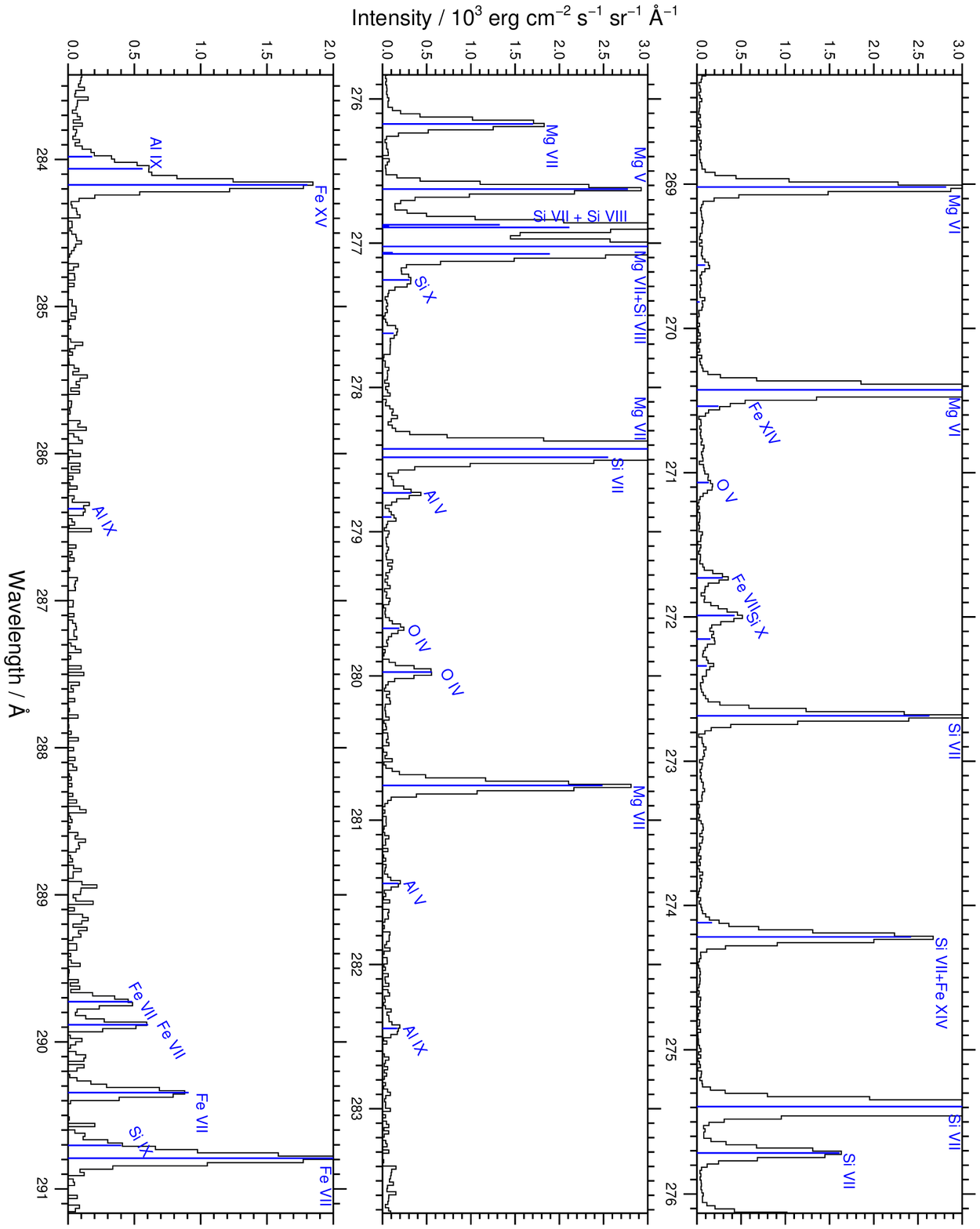}
\caption{Panels showing the long wavelength half of the EIS LW
  spectrum. See the caption of Fig.~\ref{fig.spec1} for more details.}
\label{fig.spec4}
\end{figure}

The resulting spectrum is calibrated in \ecsa\ units and has
an associated error array. The complete spectrum is displayed in
Figs.~\ref{fig.spec1}--\ref{fig.spec4}. 

As EIS does not have an internal calibration lamp, it is not possible
to derive an absolute wavelength scale for spectra obtained from the
instrument without some physical assumption about the observed
plasma. 
The emission line wavelengths given in the present atlas
are simply those measured from the spectrum output
by EIS\_MASK\_SPECTRUM and no attempt has been made to adjust them
onto a reference scale.
Sect.~\ref{sect.atlas} and the individual ion sections of
Sect.~\ref{sect.ions}  discuss the velocities derived from particular
emission lines. 
Relative wavelength comparisons for EIS emission lines are estimated
to be accurate to a level of $\pm 0.002$~\AA\  (or to within 2.1--3.5~\kms,
depending on wavelength) based on the work of \citet{brown08}.



Following creation of a single 1D spectrum for each of the EIS
wavelength bands by EIS\_MASK\_SPECTRUM, the 
Gaussian
fitting routine SPEC\_GAUSS\_EIS was used to manually derive line fit
parameters for each emission line in the spectrum. SPEC\_GAUSS\_EIS
makes use of the MPFIT procedures of 
C.~Markwadt\footnote{http://www.physics.wisc.edu/$\sim$craigm/idl/.}.
Depending on the density of lines in the spectrum, either lines were
fitted individually with single Gaussians, or multiple Gaussians were
fit simultaneously to a group of lines. In cases where good fits were
not obtained, it was sometimes necessary to force the widths of the
Gaussians to be the same. For two particularly complicated spectral
features -- the \ionx{Fe}{ix}--\ionx{Fe}{xi}--\ionx{O}{v} blend at
192.5--193.3~\AA\ (Sect.~\ref{sect.o5}), and the region around the
\ion[Si viii] lines at 276--278~\AA\ 
(Sect.~\ref{sect.si7}), a 
customized fit function was necessary to yield accurate line
parameters. The complete list of line fit parameters is given in
Table~\ref{linelist}. 


\section{Method of analysis}

The present spectrum displays a large number of lines that have either
not been identified before or not been studied in any detail. The aim
of the present work is thus threefold: (i) to use the lines to determine
the density, temperature and DEM of the plasma, (ii) confirm line 
identifications and check atomic physics properties, and (iii) build
the spectral atlas. A variety of techniques are used in the present 
work and are summarised below.

The most basic method for identifying new lines and line blends is to
study an intensity map formed from the line and compare it with a map
from an unblended line with a known temperature of formation. The 
intensity maps that we have taken as a reference are displayed in 
Figure~\ref{maps}, and the comparison allows us to associate each 
line to a temperature class. The temperature classes are listed in 
Table~\ref{classes}, together with the ions most representative of 
that class, for which we mostly chose iron ions. Note that the 
temperatures given in Table~\ref{classes} do not necessarily agree 
with the temperatures of maximum ionization derived from theoretical 
calculations \citep[e.g.,][]{bryans09} as we believe these calculations 
are not accurate for some ions. The partition of the ions in so many 
classes was made possible by the excellent signal-to-noise ratio 
provided by EIS for some of the lines of each representative ion. 
In the case of weaker lines, sometimes the small details that 
discriminate two adjacent classes were lost in the noise: in these 
cases, both classes are listed in the atlas separated by a dash 
line: for example, ``C-D'' means that this line could either 
belong to class C or to class D.

\begin{table}
\caption{Temperature classes used to classify lines in Table~\ref{linelist}. \label{classes}} 
\begin{center}
\begin{tabular}{cll}
\tableline\tableline
Class & $\log T$ & Ions \\
\tableline
A & $< 5.45$ & \ion[He ii], \ion[O iv-vi] \\
B & 5.45     & \ion[Mg v] \\
C & 5.60     & \ion[Fe vii] \\
D & 5.75     & \ion[Fe viii] \\
E & 5.90     & \ion[Fe ix] \\
F & 6.00     & \ion[Fe x] \\
G & 6.05     & \ion[Fe xi] \\
H & 6.20     & \ion[Fe xii-xiii] \\
I & 6.25     & \ion[Fe xiv] \\
L & 6.35     & \ion[Fe xv] \\
M & 6.40     & \ion[Fe xvi] \\
\tableline
\end{tabular}
\end{center}
\end{table}

Intensity map classes also helped us discriminate the cases were lines 
emitted by ions formed at much different temperatures were blended together. 
These cases could be easily identified, for example, when intensity maps 
showed features of both a cold line and a hot line. These cases have been 
marked by listing in the atlas both classes, separated by a comma: for 
example ``C,L'' marks a blend between a C and an L line.

For a number of lines in the atlas it was not possible to assign a temperature 
class, as the intensity maps were too noisy. For some weak lines the bright knot 
of emission seen in the cool lines Fig.~\ref{maps} could be clearly discerned, 
but it was not possible to clearly assign the line to a definite
class. Such lines are indicated by a spectral class of ``A--D'' since
the knot of emission is a strong feature at each of these
temperatures.


Once lines have been identified, or provisionally identified, a first
determination of the DEM of the plasma is made using the iterative
technique by Landi \& Landini (1997). The resulting curve is used 
in combination with the L-function method by Landi \& Landini (1997)
to simultaneously compare all the lines from a given ion, yielding 
density estimates when density sensitive lines are available, and 
highlighting lines discrepant with theory. The method is described 
in more detail in Sect.~\ref{sect.lfunction}.

Once a set of emission lines free of blending or atomic data problems
has been identified, then these are used to derive a final, more 
accurate DEM curve. This in turn is used to derive complete 
CHIANTI synthetic spectra for the EIS wavelength bands, allowing 
a final check of line blending and identification, and carry out 
a detailed comparison between CHIANTI emissivities and observed 
intensities for ions in the $\log\,T=5.0$--5.9 temperature range.

For calculating line intensities and computing the DEM curve, all 
atomic data were taken from CHIANTI version 6.0 \citep{dere09}. The 
DEM curve has been derived adopting the ion abundances in \cite{dere09} 
and the Feldman \etal (1992) element abundances.

\subsection{DEM measurements}\label{sect.dem}

The DEM diagnostic technique we use is described in \citet{landi97} and is
briefly summarized here. The line flux emitted by an optically thin plasma observed
at distance $d$ is given by

\begin{equation} 
F_{ij} = {1\over{4 \pi d^2}}\int{G{\left({T,N_e}\right)}\varphi{\left({T}\right)}dT} \quad\quad\quad\quad {\mathrm{ph~cm^{-2}~s^{-1}}}
\end{equation} 

\noindent
where the Contribution Function is defined as 

\begin{equation}
{G_{ij}{\left({T,N_e}\right)}} = {{N_j{\left({X^{+m}}\right)}}\over{N{\left({X^{+m} }\right)}}}~{{N{\left({X^{+m}}\right)}}\over{N{\left({X}\right)}}}~{{N{\left({X} \right)}}\over{N{\left({H}\right)}}}~{{N{\left({H}\right)}}\over{N_e}}~{A_{ij}\over {N_e}}
\end{equation}

\noindent
and the volume Differential Emission Measure (DEM) is defined as

\begin{equation}
\varphi{\left({T}\right)} = N_e^2 {{dV}\over{dT}}
\end{equation} 

\noindent
An initial, arbitrary DEM $\varphi_o(T)$ is first adopted; using a correction function
$\omega{\left({T}\right)}$, the true DEM curve is given by

\begin{equation}
\varphi{\left({T}\right)} = \omega{\left({T}\right)}~\varphi_o{\left({T}\right)}
\end{equation}

\noindent
If we define the effective temperature \tef as

\begin{equation} 
Log T_{eff} = {{\int G_{ij}{\left({T}\right)}~\varphi{\left({T}\right)}~logT~dT}\over{\int
G_{ij}{\left({T}\right) }~\varphi{\left({T}\right)}~dT}}
\end{equation}

\noindent
it can be easily shown that, as long as the correction function is slowly varying,

\begin{equation}
\label{citami}
I_{ij}={1 \over{4 \pi}}~\omega{\left({T_{eff}}\right)}~\int~G_{ij}{\left({T}\right)}~\varphi_o{\left({T}\right)}~dT
\end{equation}

From equation~\ref{citami}, each observed line flux can be used to determine the
correction function at temperature \tef; if lines from many ions are available
the $\omega{\left({T}\right)}$ curve can be sampled at many temperatures, interpolated,
and used to calculate $\varphi{\left({T}\right)}$. The resulting $\varphi{\left({T}\right)}$
curve is taken as the new trial DEM. Then the procedure is repeated until either the 
$\omega(T_{eff})$ are all equal to 1 within the errors, or the best $\chi^2$ is reached.

\subsection{L-function method of analysis}\label{sect.lfunction}

We used the temperature and density diagnostic procedure that was first introduced
by \citet{landi97}. This technique relies 
on the fact that the ${G_{ij}{\left({T,N_e}\right)}}$ curve can be expressed as

\begin{equation}
{G_{ij}{\left({T,N_e}\right)}} = f_{ij}{\left({N_e,T}\right)}~g{\left({T}\right)}
\end{equation}

\noindent
where $g{\left({T}\right)}$ is the ion abundance, it is function of temperature alone, 
and it is identical for all the lines of the same ion; while $f_{ij}{\left({N_e,T}\right)}$ 
is the population of the upper level and can be approximated with a linear function of 
$\log T$ in the temperature range where the line is formed. \citet{landi97} showed that 
an effective temperature $T_{eff}$ can be defined as

\begin{equation}
Log T_{eff} = {{\int g{\left({T}\right)}~\varphi {\left({T}\right)}~logT~dT}\over {\int
g{\left({T}\right)}~\varphi{\left({T}\right)}~dT}}
\label{teff}
\end{equation}

\noindent
and used to calculate the effective emission measure $L_{ij}{\left({N_e}\right)}$
(L-function) as

\begin{equation}
L_{ij}{\left({N_e}\right)} = {I_{obs}\over {G_{ij}(T_{eff},N_e)}}
\end{equation}

\begin{figure}
\includegraphics[width=12.0cm,height=15.0cm,angle=90]{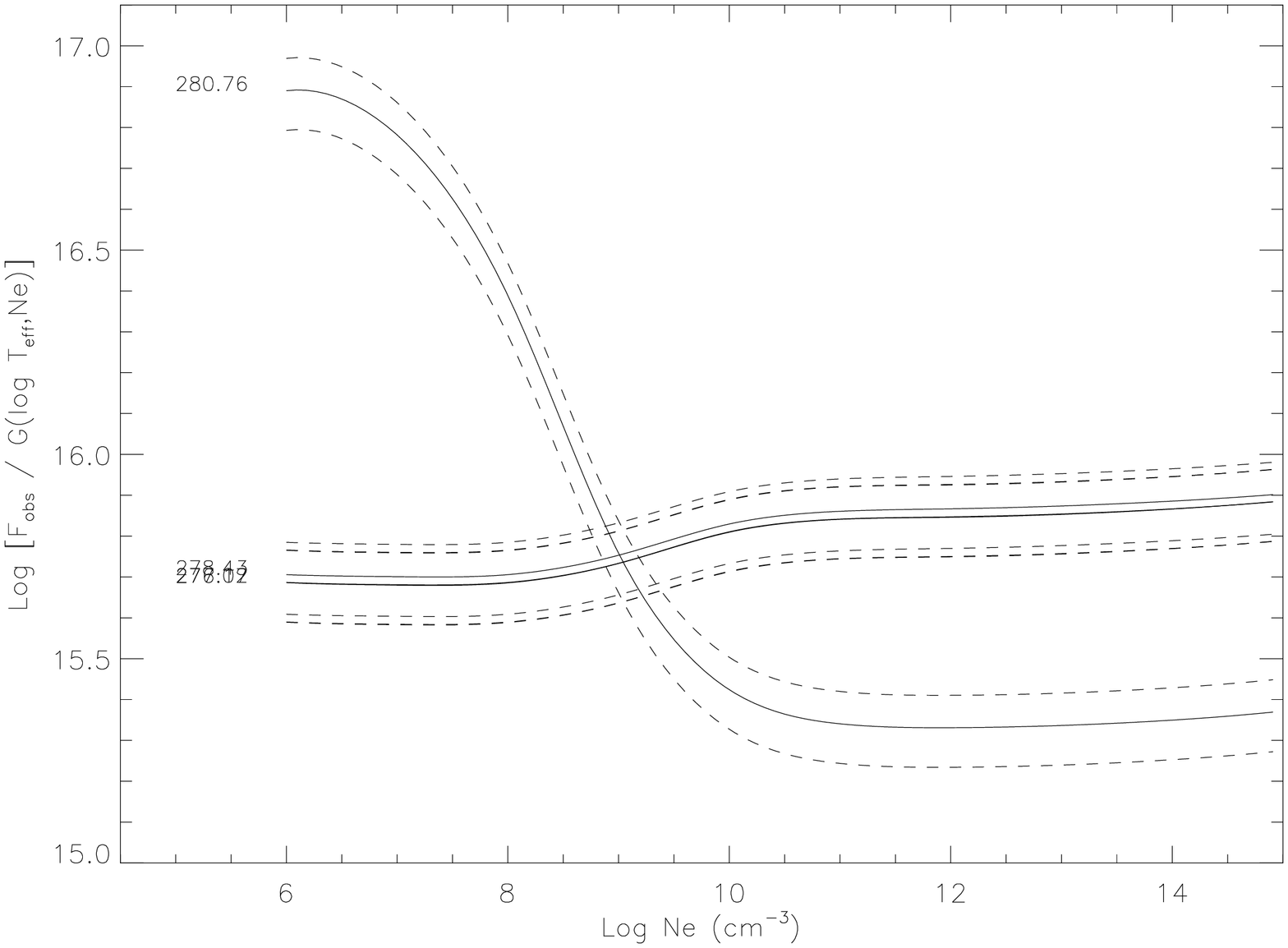}
\caption{\label{lfunction} L-function curves of the \ion[Mg vii] lines, with a 20\% uncertainty
(dotted lines). Two \ion[Mg vii] lines provide coincident L-functions.}
\end{figure}

\noindent
If we plot all the L-functions 
measured for the same ion versus the electron density, all the curves should meet in 
a common point ($N_e^{\star},L{\left({N_e^{\star}}\right)}$); the L-functions of 
density independent lines are overlapping and they also cross the same point as the 
others. An example is shown in Figure~\ref{lfunction}. Landi \& Landini (1997) showed
that the abscissa $N_e^{\star}$ of the common point is the density of the emitting
plasma.

\section{L-function/DEM results}

\begin{figure}
\includegraphics[width=12.0cm,height=15.0cm,angle=90]{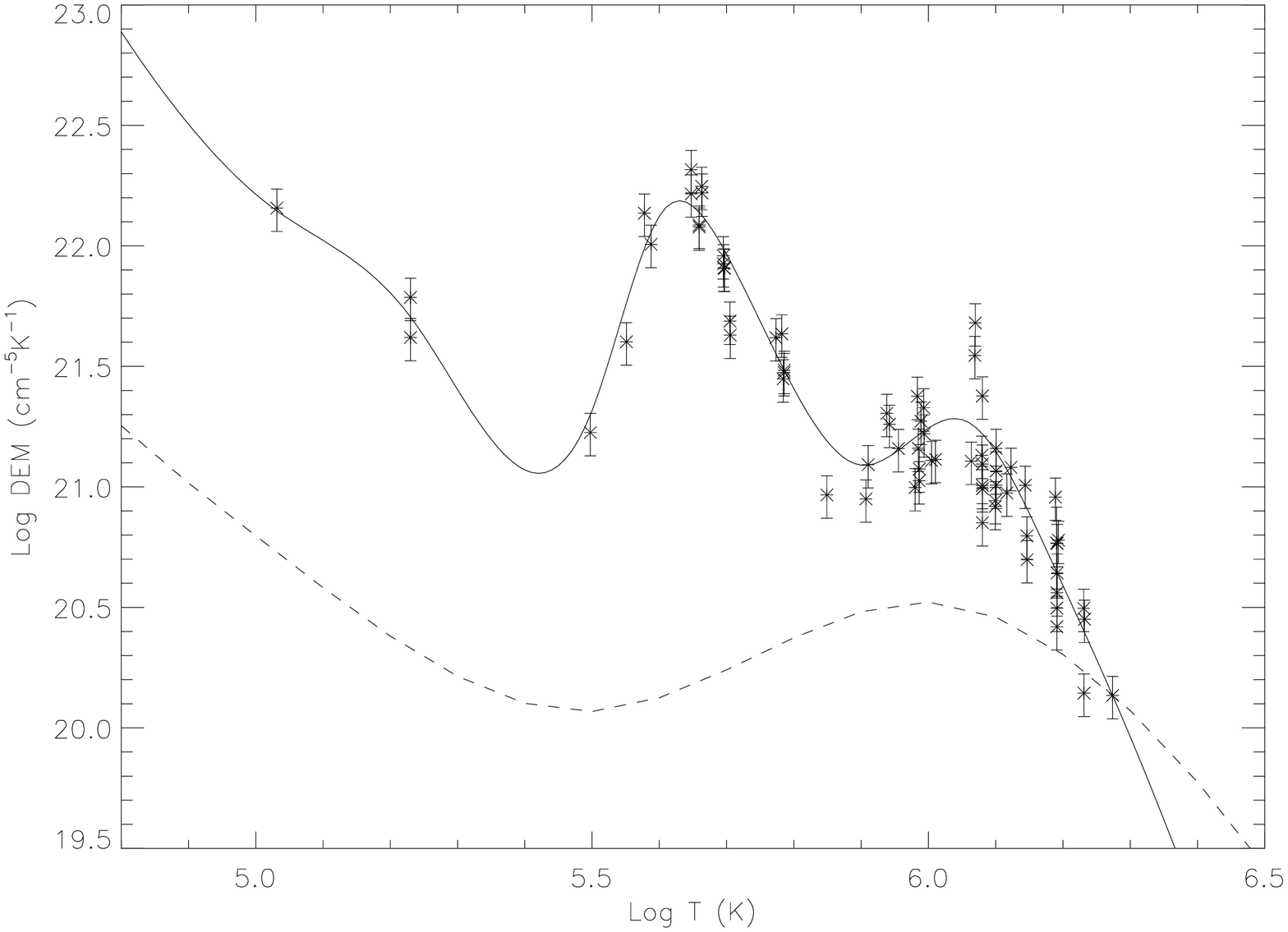}
\caption{\label{dem_first}DEM curve obtained before using the L-function method.
The CHIANTI quiet Sun DEM is superimposed for comparison purposes.}
\end{figure}

The DEM diagnostic technique described in Sect.~\ref{sect.dem} was applied 
first to a set of bright emission lines selected from a wide range of ions. 
The resulting DEM curve is displayed in Figure~\ref{dem_first} where it is 
seen that the emitting region is characterized by a very large and rather 
broad maximum at around $\log T=5.7$ that dominates the DEM at all 
temperatures larger than $\log T=5.0$. The colder regions of the DEM 
are similar to the DEM curves available in the literature (and shown in 
Figure~\ref{dem_first}, taken from the DEM quiet Sun curve available in 
the CHIANTI database), but they are much larger. Only in the corona the 
present DEM shows the same values as the CHIANTI quiet Sun DEM.

With the first-guess DEM defined, the L-functions could be calculated for 
each line (Sect.~\ref{sect.lfunction}). For each of the ions \ion[O iv-vi], 
\ion[Mg vi-vii], \ion[Al v,viii,ix], \ion[Si vi-x], \ion[S viii,x], and
\ion[Fe viii-xiv] more than one emission line is available in the spectrum 
and so the L-function method was applied to check for discrepancies and 
identify density diagnostics. A summary of derived densities is given in 
Table~\ref{density}. Details of the results of the L-function technique
are given in Table~\ref{lfunction.results}.

\begin{figure}
\includegraphics[width=12.0cm,height=15.0cm,angle=90]{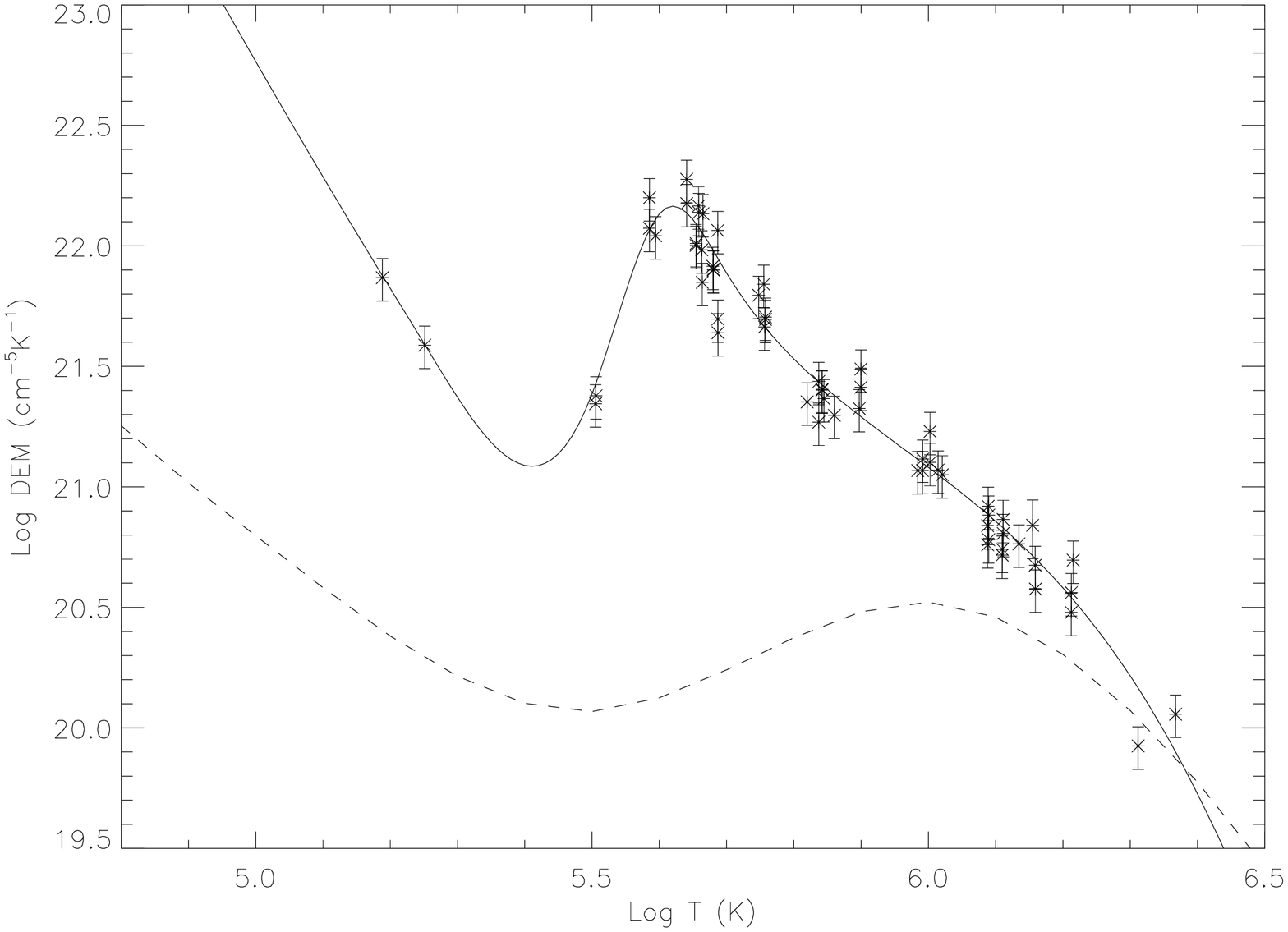}
\caption{\label{dem_second} DEM curve obtained with the spectral lines selected 
using the L-function method. The CHIANTI quiet Sun DEM is superimposed for 
comparison purposes. The lines used for generating this DEM curve had been 
previously selected with the L-function method and the first-cut DEM curve 
in Figure~\ref{dem_first}.}
\end{figure}

Filtering out emission lines that are clearly discrepant by the L-function 
method, a new DEM is calculated and the results are shown in Fig.~\ref{dem_second}. 
For this DEM we have also added the \ion[Fe xv] 284.16~\AA\ and \ion[Fe xvi] 
262.98~\AA\ lines to constrain the high temperature part of the DEM curve. 
Although these latter lines could not be tested using the L-function method 
as they are the only lines detected from those two ions in the present spectrum, 
they have been shown in the past to be free of problems (Young \etal 1998) and 
so their use does not introduce any additional uncertainty.

In computing the second DEM curve, a constant density of $\log
N_e=9.15$ was assumed. This value was selected as best value based on
the density measurements from each individual ion listed in Table~\ref{density}.

The final DEM is quite different from standard solar DEM curves,
exhibiting a large maximum at around $\log T=5.65$ that is responsible
for the strongly enhanced lines from the upper transition region ions.
A secondary maximum, corresponding to the temperature of the maximum 
of standard quiet Sun DEMs (also shown in the figure), is located just 
above $\log T=6.0$. Some plasma at active region temperatures is also 
present along the line of sight, but its importance is limited as the 
DEM curve decreases very rapidly beyond the maximum at $\log T=6.0$. 
At nearly all temperatures, the final DEM is larger than the standard
quiet Sun DEM from the CHIANTI database.

\begin{table}
\caption{\label{density} Electron density diagnostics for the selected region. $T_{eff}$ is
the effective temperature of the emitting plasma, as defined in Equation~\ref{teff}.}
\begin{center}
\begin{tabular}{lcl}
\tableline
\tableline
Ion & $\log T_{eff}$ (K) & $\log N_e$ (cm$^{-3}$) \\
\tableline
\ion[O v]     & 5.50 & $< 10.5$ \\
\ion[Mg vi]   & 5.68 & $> 7.1$ \\
\ion[Si vii]  & 5.76 & $> 7.5$ \\
\ion[Cr viii] & 5.77 & 9.45$\pm$0.30 \\
\ion[Mg vii]  & 5.78 & 9.05$\pm$0.30 \\
\ion[Fe ix]   & 5.81 & $> 8.8$ \\
\ion[Fe x]    & 5.92 & 9.00$\pm$0.25 \\
\ion[Si viii] & 5.90 & 9.05$\pm$0.30 \\
\ion[Si ix]   & 5.99 & 9.20$\pm$0.30 \\
\ion[Fe xi]   & 6.01 & 9.35$\pm$0.25 \\
\ion[Si x]    & 6.08 & 8.8$\pm$0.7 \\
\ion[Fe xii]  & 6.09 & 9.50$\pm$0.25 \\
\ion[Fe xiii] & 6.14 & 9.0$\pm$0.2 \\
\tableline
\end{tabular}
\end{center}
\end{table}

\section{Atlas}\label{sect.atlas}

The DEM curve shown in Figure~\ref{dem_second} was used to calculate a synthetic
spectrum that was used to help in the identification of the lines measured in the 
observed spectrum. Table~\ref{linelist} presents the Gaussian line fit parameters 
for every emission line in the spectrum, together with line identifications and 
predicted line intensities for the identified transitions. The coolest ion line 
in the spectrum is \ion[He ii] Lyman-$\beta$, formed at around 80,000~K, and the 
hottest line is \ionx{Fe}{xvi} \lam262.98, formed at around 2.5~million~K. Many 
of the line identifications were given in the spectral atlas of \citet{brown08}, 
but some typographical errors in that work have been corrected and, additionally, 
the DEM analysis has demonstrated that some of the proposed lines of \citet{brown08} 
do not make any significant contribution to the present spectrum. In total there 
are 277 emission lines listed in Table~\ref{linelist}, 103 of which are unidentified.


The EIS dispersion relation was derived by \citet{brown07} using mainly strong
emission lines from the iron ions \ionx{Fe}{ix--xvi} and, in particular, none 
of the cool oxygen, magnesium and silicon species discussed in the following  
sections were used. Since these lines are very strong in the present spectrum, 
we can use the measured wavelengths to investigate the accuracy of the reference 
wavelengths for these ions. Table~\ref{tbl.vel} compares velocities derived 
using the reference wavelengths given in Table~\ref{linelist} and reference 
wavelengths given in version~3 of the online NIST Atomic Database. The 
Table~\ref{linelist} wavelengths are mainly from the work of B.~Edl\'en 
\citep{edlen79, edlen83,edlen84,edlen85}. Individual ions are discussed 
in Section~\ref{sect.ions}, but we note that the Edl\'en wavelengths 
generally give a more consistent set of velocities than the wavelengths 
obtained from the NIST database. In particular, the strong lines of 
\ionx{Mg}{vii} and \ionx{Si}{vii} which are formed at a very similar 
temperature, show much better agreement with the Edl\'en wavelengths.

Comparing the velocities in Table~\ref{tbl.vel} reveals that lines
formed between $\log\,T=5.2$ to 5.8 generally have a consistent
velocity of around $+40$~\kms, while the slightly hotter lines formed
between $\log\,T=5.8$ to 6.0 are around $+20$~\kms. Inspection of the
images in Fig.~\ref{maps} shows that the bright knot of emission from
which the spectrum is obtained is most clearly visible from \ionx{O}{v}
to \ionx{Fe}{ix}, corresponding to temperatures $\log\,T=5.3$ to 5.8. The
change in line velocities is thus likely due to an increasing
contamination of the bright point spectrum by other active region
emission in the line of sight which is at a different velocity.

\section{Individual ion details}
\label{sect.ions}

In the sections below we discuss identifications and diagnostics,
and compare CHIANTI emissivities and observed line intensities for 
species found in the spectrum. The discussion is focussed towards 
ions with $\log\,T_{\rm eff} \leq 6.0$ as these lines are more 
intense in this spectrum than other types of solar spectra. For 
ions outside this temperature range, particularly the iron ions 
\ion[Fe xi-xiv], a detailed study of line ratios and identifications 
is deferred to a future paper. The results of the L-function 
diagnostics are given in Table~\ref{lfunction.results}, where we
provide the L-function values of all the lines and their ratios
to the lowest one, calculated at the crossing point. If all lines
are density insensitive relative to each other, or the L-functions
do not provide a clear indication of $\log N_e$, the ratios were
calculated assuming $\log N_e=9.15$. Each line intensity has been 
associated to an uncertainty of 20\%, to account for atomic data
uncertainties.

\begin{deluxetable}{ccrl}
\tabletypesize{\footnotesize}
\tablecaption{L-function results.\label{lfunction.results}}
\tablewidth{0pt}
\tablehead{\colhead{$\lambda$} & \colhead{$L{\left({T_{eff}}\right)}$} & \colhead{Ratio} & \colhead{Notes}}
\tablecolumns{4}
\startdata
\ion[O iv]    & $\log T_{eff} = 5.19$ & & \\
279.67 & 15.09$\pm$0.09 &   0.69 & NO \\
279.97 & 15.25$\pm$0.09 &   1.00 & OK \\
260.30 & 15.27$\pm$0.09 &   1.04 & OK \\
 & & & \\
\hline
 & & & \\
\ion[O v]     & $\log T_{eff} = 5.50$ & & \\
185.78 & 15.55$\pm$0.12 &   3.31 & Unidentified blend; \ion[O v] $\simeq 30$\%   \\
192.81 & 15.46$\pm$0.09 &   6.99 & Blended with \ion[Fe xi]; \ion[O v] $\simeq 37$\%   \\
192.93 & 15.03$\pm$0.09 &   1.00 & Blended with \ion[Fe xi]; \ion[O v] $\simeq 100$\% --OK  \\
248.49 & 15.11$\pm$0.09 &   1.20 & Blended with \ion[Al viii]; \ion[O v] $\simeq 83$\% -- OK  \\
271.07 & 15.40$\pm$0.09 &   2.34 & Blended with \ion[Fe vii]; \ion[O v] $\simeq 43$\%   \\
 & & & \\
\hline
 & & & \\
\ion[O vi]    & $\log T_{eff} = 5.65$ & & \\
172.93 & 16.09$^{+0.19}_{-0.34}$ & 1.74 & OK \\
173.12 & 15.97$^{+0.15}_{-0.24}$ & 1.32 & OK \\
183.95 & 15.98$\pm$0.09 &   1.35 & Unidentified blend; \ion[O vi] $\simeq 74$\%  \\
184.14 & 15.85$\pm$0.09 &   1.00 & OK   \\
 & & & \\
\hline
 & & & \\
\ion[Mg vi]   & $\log T_{eff} = 5.66$ & & \\
269.02 & 15.76$\pm$0.09 &   1.00 & OK \\
270.43 & 15.80$\pm$0.09 &   1.10 & OK \\
 & & & \\
\hline
 & & & \\
\ion[Mg vii]  & $\log T_{eff} = 5.76$ & & \\
276.17 & 15.73$\pm$0.09 &   1.00 & OK \\
277.04 & 15.73$\pm$0.09 &   1.00 & OK \\
278.44 & 15.75$\pm$0.09 &   1.05 & OK \\
280.76 & 15.77$\pm$0.09 &   1.10 & OK \\
 & & & \\
\hline
 & & & \\
\ion[Al v]    & $\log T_{eff} = 5.56$ & & \\
278.73 & 15.74$\pm$0.09 &   1.35 & \ion[Ni xi]; \ion[Al v] $\simeq 74$\% -- OK  \\
281.44 & 15.61$\pm$0.09 &   1.00 & OK \\
 & & & \\
\hline
 & & & \\
\ion[Al vii]  & $\log T_{eff} = 5.76$ & & \\
259.23 & 16.33$\pm$0.09 &   5.89 & Blend \ion[Cr vii]; \ion[Al vii] $\simeq 17$\% -- NO   \\
261.24 & 15.56$\pm$0.09 &   1.00 & OK   \\
 & & & \\
\hline
 & & & \\
\ion[Al viii] & $\log T_{eff} = 5.90$ & & \\
247.43 & 16.20$\pm$0.09 &   6.61 & Unidentified blend; \ion[Al viii] $\simeq 15$\%   \\
248.49 & 16.23$\pm$0.09 &   7.08 & Blended with \ion[O v]; \ion[Al viii] $\simeq 15$\% -- OK  \\
250.16 & 15.38$\pm$0.09 &   1.00 & OK   \\
 & & & \\
\hline
 & & & \\
\ion[Al ix]   & $\log T_{eff} = 6.00$ & & \\
282.44 & 15.31$\pm$0.09 &   1.35 & Unidentified blend; \ion[Al ix] $\simeq 74$\%     \\
284.06 & 15.42$\pm$0.09 &   1.74 & Unidentified blend; \ion[Al ix] $\simeq 58$\%     \\
286.38 & 15.18$\pm$0.09 &   1.00 &      \\
 & & & \\
\hline
 & & & \\
\ion[Si vi]   & $\log T_{eff} = 5.66$ & & \\
246.04 & 15.64$\pm$0.09 &   1.02 & OK \\
249.16 & 15.63$\pm$0.09 &   1.00 & OK \\
 & & & \\
\hline
 & & & \\
\ion[Si vii]  & $\log T_{eff} = 5.76$ & & \\
272.69 & 15.64$\pm$0.09 &   1.10 & OK \\
274.22 & 15.79$\pm$0.09 &   1.55 & Blend \ion[Fe xiv]; \ion[Si vii] $\simeq 65$\%   \\
275.39 & 15.60$\pm$0.09 &   1.00 & OK \\
275.71 & 15.64$\pm$0.09 &   1.10 & OK \\
276.88 & 15.64$\pm$0.09 &   1.10 & OK \\
278.44 & 15.62$\pm$0.09 &   1.05 & OK \\
 & & & \\
\hline
 & & & \\
\ion[Si viii] & $\log T_{eff} = 5.90$ & & \\
250.51 & 15.70$\pm$0.09 &   1.23 & OK \\
250.82 & 15.62$\pm$0.09 &   1.02 & OK \\
276.85 & 15.63$\pm$0.09 &   1.05 & OK \\
276.87 & 15.61$\pm$0.09 &   1.00 & OK \\
277.04 & 15.63$\pm$0.09 &   1.05 & OK \\
277.06 & 15.61$\pm$0.09 &   1.00 & OK \\
 & & & \\
\hline
 & & & \\
\ion[Si ix]   & $\log T_{eff} = 6.02$ & & \\
258.09 & 15.20$\pm$0.09 &   1.05 & OK \\
290.71 & 15.18$\pm$0.09 &   1.00 & OK \\
 & & & \\
\hline
 & & & \\
\ion[S viii]  & $\log T_{eff} = 5.88$ & & \\
198.57 & 15.75$\pm$0.09 &   1.00 & Blend \ion[Fe xi]   \\
202.62 & 15.88$\pm$0.09 &   1.35 & Blend \ion[Fe xi]   \\
 & & & \\
\hline
 & & & \\
\ion[Cr vii]  & $\log T_{eff} = 5.66$ & & \\
202.86 & 15.66$\pm$0.09 &   1.00 & OK   \\
258.64 & 16.27$\pm$0.09 &   4.07 & Unidentified blend; \ion[Cr vii] $\simeq 25$\%   \\
259.23 & 16.33$\pm$0.09 &   4.67 & Blend \ion[Al vii]; \ion[Cr vii] $\simeq 21$\% -- NO  \\
261.35 & 16.68$\pm$0.09 &  10.50 & Unidentified blend; \ion[Cr vii] $\simeq 10$\%   \\
 & & & \\
\hline
 & & & \\
\ion[Cr viii] & $\log T_{eff} = 5.77$ & & \\
205.05 & 16.06$\pm$0.09 &   1.00 & OK \\
205.72 & 16.09$\pm$0.09 &   1.07 & OK    \\
208.68 & 16.11$\pm$0.09 &   1.12 & OK \\
211.48 & 16.08$\pm$0.09 &   1.05 & Blend \ion[Ni xi]; \ion[Cr viii] $\simeq 95$\% -- NO   \\
 & & & \\
\hline
 & & & \\
\ion[Mn viii] & $\log T_{eff} = 5.71$ & & \\
185.46 & 15.73$\pm$0.09 &   1.00 & OK \\
263.20 & 16.10$\pm$0.09 &   2.34 & Unidentified blend; \ion[Mn viii] $\simeq 43$\%   \\
 & & & \\
\hline
 & & & \\
\ion[Mn ix]   & $\log T_{eff} = 5.86$ & & \\
188.43 & 15.97$\pm$0.09 &   1.59 & Blend \ion[Fe vii]  \\
191.61 & 16.00$\pm$0.09 &   1.70 &  \\
199.33 & 15.77$\pm$0.09 &   1.00 &  \\
 & & & \\
\hline
 & & & \\
\ion[Ni xi]   & $\log T_{eff} = 5.95$ & & \\
207.95 & 15.26$\pm$0.09 &   1.00 & OK \\
211.48 & 15.51$\pm$0.09 &   1.78 & Blend \ion[Cr viii]; \ion[Ni xi] $\simeq 56$\% -- NO  \\
278.73 & 16.17$\pm$0.09 &   8.13 & Blend \ion[Al v]; \ion[Ni xi] $\simeq 12$\% -- OK  \\
 & & & \\
\hline
\enddata
\end{deluxetable}

In Table~\ref{lfunction.results}, if a ratio is less than 1 it indicates
the measured line intensity is weaker than predicted by theory, while a
ratio greater than 1 indicates the measured line is stronger than predicted
by theory. Usually the latter case is due to a blending line.

\subsection{\ion[O iv]}
\label{sect.o4}

A large number of $n=2$ to $n=3$ transitions lie in the wavelength
range 170--292~\AA, with the brightest two lines falling between the
two EIS wavebands at around 238.5~\AA. The next strongest is the
$2s^22p$ $^2P_{3/2}$ -- $2s^23s$ $^2S_{1/2}$ transition at
279.933~\AA, which is seen in the EIS spectrum. Another decay from
this upper level to $2s^22p$ $^2P_{1/2}$ falls nearby at 279.631~\AA\
and this is also seen in the spectrum. While the observed separation
of the two lines is in excellent agreement with the wavelengths of
\citet{edlen34}, the branching ratio of the lines shows a significant
discrepancy with the prediction from CHIANTI: the predicted ratio
being 0.50 compared to the observed ratio of $0.34\pm 0.05$. The line
widths show no indication of blends (Table~\ref{linelist}), and the
L-function method suggests the \lam279.93 yields  the best agreement
between theory and observation (Tables~\ref{lfunction.results} and
\ref{linelist}). 

A number of transitions between excited configurations are predicted
in the EIS wavelength range, the strongest of which is potentially
observable. The $2s2p^2$ $^2D_{5/2}$ -- $2s2p(^3P)3d$ $^2F_{7/2}$
transition was found at 260.389~\AA\ by \citet{edlen34}. A possible candidate is the line measured
at 260.292~\AA\ for which the intensity is in excellent agreement with
\lam279.93 (Table~\ref{lfunction.results}), however the wavelength
shows a significant discrepancy: the measured wavelength implying a
velocity of $-112$~\kms\ compared to $+43$~\kms\ for \lam279.93. The
width of the line is also significantly broader than \lam279.93.
The ratios of the observed 260.29~\AA\ line relative to either
\lam279.63 or \lam279.93 are excellent temperature diagnostics with
little sensitivity to density
and the derived values are $\log T=5.18\pm$0.05  and 5.30$\pm$0.07 for
\lam279.93 and \lam279.63, respectively.


Additional weak lines are expected to fall in the EIS wave bands, but they are less
intense than those reported in the atlas, and they are not found in
the spectrum. A line with rest wavelength 203.044~\AA\ is a good
wavelength match for 
the line observed at 203.064~\AA\, but
the observed intensity is a factor 5 larger than predicted, so \ion[O iv] only provides
a minor contribution to the observed feature.


\subsection{\ion[O v]}
\label{sect.o5}

There are several \ion[O v] lines in the EIS wavelength range, but
most are affected by blending. A few of these lines are density sensitive
relative to each other, but they only allow to determine an upper limit
to the plasma electron density, $\log N_e < 10.5$. The strongest line by 
intensity is the $2s2p$ $^1P_1$ -- $2s3s$ $^1S_0$ transition  at 248.46~\AA, 
which is blended with an \ion[Al viii] line. Table~\ref{lfunction.results}
shows that the contribution of  \ion[Al viii] to the line is $\simeq 15$\% 
in agreement with the L-function results for \ion[Al viii].

A group of six transitions from the \orb[2s2p ] \tm[3 P ] -- \orb[2s3d ]
\tm[3 D ] multiplet are found between 192.75 and 192.91~\AA\ and have
been discussed by \citet{young07b}. They are partly blended with
\ion[Fe xi] and \ion[Ca xvii] lines and a method to extract the
intensities of the individual component lines has been described by
\citet{ko09}. In the present case a slightly modified treatment is
used since there is very little high temperature emission in the
spectrum and so \ion[Ca xvii] can be safely ignored. In addition a
nearby line at 192.64~\AA, which we believe is due to \ion[Fe ix]
\citep{young09a} is quite strong and needs to be accounted for in the
fit. We include six Gaussians for the six \ion[O v] lines,
with the separations being fixed to the separations of the CHIANTI
wavelengths and the widths forced to be the same. Although there is
some density sensitivity amongst the lines, it is small and we force
the lines to have the relative strengths predicted by CHIANTI at a
density of $10^{10}$~cm$^{-3}$. In summary, then, the only free
parameters for the \ion[O v] lines are taken to be the wavelength,
width and amplitude of the \lam192.904 line (the strongest of the
group), with the parameters for the other lines all fixed relative to
this. For \ion[Fe xi] \lam192.83 and the line at 192.64~\AA\ we fit
two independent Gaussians. Note that the additional \ionx{Fe}{xi}
\lam192.90 line discussed by \citet{ko09} is not included as it is
very weak. The resulting fit parameters for the Gaussians are given 
in Table~\ref{linelist}. Note that the fit parameters for each of 
the \ion[O v] lines except \lam192.904 are derived from the \lam192.904
fit parameters as described above. Confidence in the derived fit 
parameters is obtained by comparing the velocity shifts of \lam192.83 
and \lam192.904 ($-36$~\kms\ and $+42$~\kms, respectively) with 
\ionx{Fe}{xi} \lam188.23 and \ionx{O}{v} \lam248.46 ($-35$~\kms\
and $+35$~\kms, respectively). 

Three further \ionx{O}{v} lines are predicted in the EIS wavebands. 
\lam172.17 is comparable in strength to \lam192.90 but the instrument 
effective area is much lower in this part of the spectrum. A couple
of lines are indeed barely visible at around \lam172.0-172.3 but
they are too weak to provide a reliable measurement of their parameters.
The observed line at 185.780~\AA\ is a good 
wavelength match for \ionx{O}{v} \lam185.745 however the intensity 
map for the line indicates that it is emitted by an unidentified 
hotter ion formed at temperatures closer to \ion[Mg v] and \ion[Fe vii]. 
Indeed, Table~\ref{lfunction.results} indicates that \ion[O v] provides 
only $\simeq 30$\% of the observed intensity. The other \ion[O v] line 
identified in the long wavelength section of the atlas is \lam271.068, 
which sits in a rather broad spectral feature to which \ionx{Fe}{vii}
\lam271.074 also contributes \citep{young09a}. We estimate \ion[O v] 
accounts for $\simeq 43$\% of the measured intensity.

\subsection{\ion[O vi]}
\label{sect.o6}

The $2p$ $^2P_{1/2,3/2}$ -- $3s$ $^2S_{1/2}$ transitions, \lam\lam183.94, 
184.12, are seen in the present spectrum and their intensities are reproduced 
reasonably well by the DEM (Table~\ref{linelist}). However their L-functions 
show a small, but significant, discrepancy with theory (Table~\ref{lfunction.results}), 
with \lam183.94 observed to be too strong compared to \lam184.12 by a factor
1.35. This is surprising for such a simple ion and the obvious solution is 
that \lam183.94 is blended.  However, images formed in both lines are very 
similar and show no evidence of a contribution from a line formed at a 
different temperature. In terms of line widths, \lam183.94 is actually 
found to be a little narrower than \lam184.12 and so any blending line 
must lie at almost exactly the same wavelength as \lam183.94. Comparing 
the measured wavelengths with the rest wavelengths of \citet{edlen79} 
and converting to velocity units gives $+19.6\pm 3.3$ and $+35.8\pm 
3.3$~\kms\ for \lam183.94 and \lam184.12, respectively. The separation 
of the lines is thus not consistent with their rest wavelengths. The 
velocity of \lam184.12 is more consistent with other ions formed at a 
similar temperature (Table~\ref{tbl.vel}), suggesting problems with the 
\lam183.94 line. A study of \lam183.94 and \lam184.12 in a range of
different conditions would be valuable for further investigating these
problems.

The only other \ion[O vi] lines expected in the EIS spectrum are the
three transitions of the $2p$ $^2P_{J}$ -- $3d$ $^2D_{J^\prime}$
multiplet. The strongest line (3/2--5/2) has a rest wavelength of
173.080~\AA\ and is partly blended with the 3/2--3/2 transition at
173.095~\AA\, although the latter is predicted to be a factor 0.17
smaller. The 1/2--3/2 transition is at 172.936~\AA. The EIS effective
area is very low at these wavelengths, but two lines can be seen close
to these wavelengths (Table~\ref{linelist}). Converting the measured
wavelengths to velocity units gives $-17\pm 25$ and $69\pm 29$~\kms\
for \lam172.94 and \lam173.08, respectively, and so only the latter
line is consistent with the velocities of \lam183.94 and \lam184.12
presented above. The intensities of these lines are a little larger
than expected relative to the \lam184.14 (Table~\ref{lfunction.results}),
but the uncertainties are larger than the difference.

\subsection{\ion[Ne v]}
\label{sect.ne5}

A number of weak \ion[Ne v] transitions are predicted through the EIS
wavebands and the strongest in terms of counts expected on the
detector is \lam184.735 which is a possible wavelength match for an
observed line at 184.777~\AA\ although the implied velocity of
$+68$~\kms\ is larger than found for \ionx{O}{v}
(Table~\ref{tbl.vel}) which is formed at a
similar temperature. The intensity predicted from the DEM shows that
the \ion[Ne v] line cannot fully account for the 184.777~\AA\ line's
intensity and other contributions come from \ionx{Fe}{vii} and
\ionx{Fe}{xi} (Table~\ref{linelist}).
A \ion[Ne v] line at 173.932~\AA\ is predicted to be stronger than  
\lam184.735 by a factor three, but  the EIS sensitivity is low at 
this wavelength and the line can not be seen. No trace is found of 
the other \ion[Ne v] lines predicted by CHIANTI, with the only 
exception of \lam274.090, which lies close to an observed and 
unidentified line at 274.119~\AA, whose intensity image is consistent 
with a cool line. However, we hesitate to identify this line as
\ion[Ne v] since the predicted intensity is factor 4 lower than 
observed, and other \ion[Ne v] lines predicted to be brighter are 
not observed.

\subsection{\ion[Ne vi]}
\label{sect.ne6}

The only \ion[Ne vi] lines predicted to be observable in the 
EIS wavelength range belong to the \orb[2s2p 2] \tm[2 S 1/2] -- 
\orb[2s 2]\orb[2p ] \tm[2 P 1/2,3/2] doublet at 185.056 and 
184.945~\AA, respectively. The strongest of the two lines is 
close to the observed line at 184.922~\AA, however the velocity 
of $-37$~\kms\ is discrepant with the \ion[O vi] \lam184.12 
velocity of $+39$~\kms. The intensity prediction from the DEM 
analysis (Table~\ref{linelist}) shows that \ionx{Ne}{vi} can 
only account for $\simeq 50$\% of the observed intensity, and
the remaining contribution is due to \ionx{Fe}{vii} \lam184.886.
The \ion[Ne vi] \lam185.056 line is predicted to be around half 
the strength of \lam184.945, but it is not found in the spectrum.

\subsection{\ion[Mg v]} 
\label{sect.mg5}

\ion[Mg v] provides one strong line in the EIS spectrum, given by the 
allowed \orb[2s 2]\orb[2p 4] \tm[1 D ] - \orb[2s2p 5] \tm[1 P ] transition
observed at 276.625~\AA.
This line is very prominent in the present dataset, as \ion[Mg v] is formed
at temperatures close to the peak temperature of the DEM curve and the
intensity predicted from the DEM is very close to the measured
intensity (Table~\ref{linelist}). Using
the rest wavelength of \citet{edlen83} gives a velocity of $49.9\pm
0.05$~\kms\ which is around 10~\kms\ larger than the average velocity
of the lines formed below $\log\,T=5.8$, suggesting the rest
wavelength may be slightly in error.  
No other \ion[Mg v] line is identified in the atlas, because the few
other lines available in the EIS range are much weaker than \lam276.625.
Only two lines are potentially observable: they are weak, but are predicted
to be at $\simeq$197~\AA, where the EIS coating reflectivity is high. However,
their wavelengths are based on calculated rather than laboratory level
energies, so that it is difficult to associate them with any spectral line
with certainty. Table~\ref{classes} lists \ion[Mg v] as class~B, but the
few unidentified lines listed as class~B or even C are brighter by
an order of magnitude than the predicted \ion[Mg v] lines.

\subsection{\ion[Mg vi]} 
\label{sect.mg6}

The $2s^22p^3$ $^2D_J$ -- $2s2p^4$ $^2P_{J^\prime}$ multiplet gives
rise to two strong lines in the present spectrum: the 3/2--1/2
transition at 269.020~\AA\ and the 5/2--3/2 transition at
270.426~\AA. The latter is blended with the 3/2--3/2 component which
is predicted by CHIANTI to be 13~\%\ of the 5/2--3/2
component, and the fitted Gaussian at 270.426~\AA\ includes both
components. A further blend is with \ionx{Fe}{xiv} \lam270.52 which 
normally dominates in active region conditions. In the present
spectrum, however, the \ionx{Fe}{xiv} component is weak and can be fit
with a separate Gaussian . Table~\ref{lfunction.results} demonstrates that the two
observed lines are in good agreement with each other, however the DEM
over-predicts the strength of both lines by around 20~\%\
(Table~\ref{linelist}). Table~\ref{tbl.vel} shows that the velocities
derived using the reference wavelengths of \citet{edlen84} are
consistent with the other cool species in the spectrum. Note that the
slightly larger velocity for the 270.426~\AA\ line is likely due to the
weaker blending line which has a slightly longer rest wavelength.

The CHIANTI \ionx{Mg}{vi} model predicts many $n=2$ to $n=3$ transitions in the EIS
wavebands, but all are very weak and can not be observed in the
present spectrum. 

\subsection{\ion[Mg vii]}
\label{sect.mg7}

Only four lines of significant strength are expected in the EIS
wavebands, and each is bright in the current spectrum. The three
members of the $2s^22p^2$ $^3P_J$ -- $2s2p^3$ $^3S_1$ multiplet are
expected at 276.14, 276.99 and 278.39~\AA, but only the weakest line,
\lam276.14, is unblended. \lam276.99 is blended with \ionx{Si}{viii} 
and a method for extracting the line intensities in this difficult 
part of the spectrum is described in Sect.~\ref{sect.si8}. Although 
\lam276.99 is listed in Table~\ref{linelist} we note that the
parameters were completely determined from the \lam276.14 parameters
and so this is not an independent measurement. \lam278.39 is blended
with \ionx{Si}{vii} \lam278.45 but by fitting two Gaussians forced to
have the same width, the two lines' intensities can be extracted. The 
$2s^22p^2$ $^1D_2$ -- $2s2p^3$ $^1P_1$ transition is found at 280.72~\AA, 
and forms an excellent density diagnostic with any of the $2s^22p^2$ 
$^3P_J$ -- $2s2p^3$ $^3S_1$ multiplet.

Table~\ref{lfunction.results} shows that \lam276.14, \lam278.39 and
\lam280.72 are in good agreement with each other, however the DEM
underpredicts the lines' intensities by around 40~\%. Using the
reference wavelengths of \citet{edlen85} yields velocities for 
\lam276.14, \lam278.39 and \lam280.72 that are in good agreement 
with the other ion species formed below $\log\,T=5.8$. However, 
we note that using the reference wavelengths from the NIST database 
gives significantly lower velocities. The best agreement among 
\ion[Mg vii] lines is found at $\log N_e = 9.05\pm 0.30$.

\subsection{\ion[Al v]}
\label{sect.al5}

The \orb[2s 2]\orb[2p 5] $^2P_{3/2,1/2}$ -- \orb[2s2p 6] $^2S_{1/2}$ transitions
at 278.69 and 281.39~\AA, respectively, are the only \ion[Al v] lines visible in the
EIS wavelength band. They are emitted by the same upper level so they can not be used 
for temperature or density diagnostics. These two lines disagree with each other,
because the 278.73 line is affected by a relatively weak blend due to a previously 
unidentified \ion[Ni xi] line, expected to account for $\simeq 25$\% of the observed 
intensity. The predicted \ion[Ni xi] line intensity is a bit lower than needed, but 
the combined intensity of the two lines is reasonably close to the observed value.
The \lam281.438 line is very close to a \ion[S xi] feature prominent in active region 
plasmas, but in the present dataset the contribution of this line is negligible.

The velocities of the two \ionx{Al}{v} lines are consistent with other
species (Table~\ref{tbl.vel}) when using the reference wavelengths of
\citet{artru74}, and the DEM predictions for the two lines are in good
agreement with measurements.

\subsection{\ion[Al vii]}
\label{sect.al7}

Four transitions of the  $2s^22p^3$ $^2P_J$ -- $2s2p^4$ $^2P_{J^\prime}$ 
multiplet are predicted to lie between 259 and 262~\AA. The strongest is 
\lam261.208 which is found in the spectrum; the velocity is consistent 
with \ionx{Mg}{vii} which is formed at the same temperature (Table~\ref{tbl.vel}) 
and the DEM predicts the strength of the line to be in excellent agreement 
with observations. Another line from the same upper
level, \lam261.030, is blended with \ionx{Si}{x} \lam261.05 but we
estimate that it makes less than a 3~\%\ contribution to this line.
The $^2P_{1/2}$ level gives rise to two lines at 259.020 and
259.196~\AA, the latter of which is a good wavelength match for the
observed line at 259.226~\AA. However, Table~\ref{lfunction.results}
shows that the observed line is much stronger than expected. There is
a known blend with a \ionx{Cr}{vii} line, but this accounts for only
$\simeq 20$~\% of the total observed  
intensity, with \ion[Al vii] accounting for $\simeq 15$~\%.  
\ionx{Al}{vii} \lam259.020 is predicted to be 83~\%\ of the strength of
\lam259.196, but no line can be measured at this wavelength,
consistent with \lam259.196 providing only a small contribution to the
measured line at 259.226~\AA.

The  $2s^22p^3$ $^2P_{1/2,3/2}$ -- $2s2p^4$ $^2S_{1/2}$ transitions at
278.960 and 279.164~\AA\ are density sensitive relative to the lines
discussed above, but can not be found in the present spectrum. This is
consistent with a density of $\le 10^{10}$~cm$^{-3}$.

\subsection{\ion[Al viii]}
\label{sect.al8}

\ionx{Al}{viii} is isoelectronic with \ionx{Mg}{vii} and the four strong 
\ionx{Mg}{vii} transitions between 276 and 281~\AA\ discussed earlier are 
found between 247 and 252~\AA\ for \ionx{Al}{viii}, although much weaker 
due to the lower element abundance. The strongest of the $2s^22p^2$ $^3P_J$ 
-- $2s2p^3$ $^3S_1$ multiplet is present in the EIS spectrum at 250.155~\AA:
the line velocity is in good agreement with the \ionx{Si}{viii} lines 
which have a similar $T_{\rm eff}$ value (Table~\ref{tbl.vel}) and the
intensity is well reproduced by the DEM. The next 
strongest line is blended with \ionx{O}{v} \lam248.46 and 
Table~\ref{lfunction.results} shows that \ionx{Al}{viii} contributes 
$\simeq 15$\% of the total intensity. The third and weakest line of 
the multiplet is blended with an unknown line at 247.426~\AA, which is
expected to provide $\simeq 85$\% of the total intensity.

The density sensitive $2s^22p^2$ $^1D_2$ -- $2s2p^3$ $^1P_1$ transition 
is expected at 251.36~\AA\ but can not be found in the spectrum. Another 
density sensitive line is $2s^22p^2$ $^1D_2$ -- $2s2p^3$ $^1D_2$ at 
285.46~\AA\ which is around a factor two stronger than \lam251.36 but 
it also can not be found in the spectrum. This is consistent with an
electron density lower than $\log N_e < 10$.

\subsection{\ion[Al ix]}
\label{sect.al9}

The only \ion[Al ix] transitions which provide observable lines in the EIS ranges
are those from the \orb[2s 2]\orb[2p ] \tm[2 P ] -- \orb[2s2p 2] \tm[2 P ] multiplet,
observed in the 280-287~\AA\ wavelength range. The strongest of these lines
(\tm[2 P 3/2]-\tm[2 P 3/2]) lies close to the very strong \ion[Fe xv] line at
\lam284.15 and in active region conditions it is lost under the profile of the
latter line. The \tm[2 P ]-\tm[2 S ] transitions are found just outside the EIS
wavelength range at around 300~\AA. These lines provide density sensitive
intensity ratios in the $7.0 < \log N_e < 8.5$ range.

Three out of four \ion[Al ix] transitions are identified in the present spectrum, 
since the intensity of the weakest line of the multiplet is predicted to be
4.4~erg~cm$^{-2}$s$^{-1}$sr$^{-1}$ and is not observed. Table~\ref{lfunction.results}
shows that the other three lines are only in partial agreement with each other.
There is a factor $\simeq$1.7 difference between the $L{\left({T_{eff}}\right)}$
values of the \lam284.06 and the \lam286.38 lines which can not be accounted for
by the uncertainties. The third line falls in between these two so that it is
not easy to understand whether the strongest line of the multiplet is blended,
or some problem affects the \lam286.38 line. Density sensitivity is not the
cause of the problem, as the $L{\left({T_{eff}}\right)}$ values of the three
lines are closest to each other for $\log N_e > 8.2$, and diverge at lower
densities. The \ion[Fe xv] line is moderately weak at the locations we have 
selected for the present atlas, so it should be well resolved from the \ion[Al ix]
transition.

\subsection{\ion[Si vi]}
\label{sect.si6}

The atomic model for \ion[Si vi] in CHIANTI predicts only two bright 
emission lines in the EUV, both of which are observed by EIS. A large 
number of additional lines are predicted to be three or more orders 
of magnitude less intense and are too weak to be observed by EIS. 
The strongest line is at 246.01~\AA\ and we perform a simultaneous 
three Gaussian fit here in order to pick out two weak lines in the 
wings of the \ion[Si vi] line. The longer wavelength line is due to 
\ion[Fe xiii] while the short wavelength line is unknown. Note 
that the widths of each of the three Gaussians were forced to be 
the same in the fitting process, and thus the \ion[Si vi] width 
will dominate, making the line fit parameters for the two weak 
lines uncertain.

\ion[Si vi] \lam249.12 is close to the hot \ion[Ni xvii] \lam249.18
line which is very strong in the cores of active regions, but can be
neglected in the present spectrum. Another nearby line we believe is
due to \ion[Fe vii] but is clearly separated from the \ion[Si vi] line. 
The two \ion[Si vi] lines form a branching ratio and the predicted
intensities are in excellent agreement with the measured values 
(Table~\ref{lfunction.results}).

\subsection{\ion[Si vii]}
\label{sect.si7}

The only significant lines predicted by CHIANTI in the EIS wavebands
belong to the $2s^22p^4$ $^3P_J$ -- $2s2p^5$ $^3P_{J^\prime}$
multiplet, which yields six lines between 272 and 279~\AA. Three of
these lines are blended but in the case of \lam278.45 a two Gaussian
fit can be used to separate the line from \ionx{Mg}{vii} \lam278.39 if
both lines are forced to have the same width (Table~\ref{linelist}).
There are three lines that are emitted by the $2s2p^5$ $^3P_1$ level 
(\lam\lam272.65, 275.68, 276.85): \lam276.85 is blended with two 
\ionx{Si}{viii} transitions and Sect.~\ref{sect.si8} demonstrates 
how \ionx{Si}{vii} \lam275.68 was used to estimate the \ionx{Si}{vii} 
contribution to this blend. The two unblended lines, \lam272.65 and 
275.68, are in good agreement with theory (Table~\ref{lfunction.results}).
The three lines from the $^3P_1$ upper level are weakly density sensitive 
relative to those from the $2s2p^5$ $^3P_2$ level (\lam\lam275.35, 278.44). 
Agreement is found for any density $\log N_e \geq 7.5$.

The remaining \ion[Si vii] line at 274.18~\AA\ is emitted from 
the $2s2p^5$ $^3P_0$ level and shows greater density sensitivity 
than the other lines. It is blended with \ion[Fe xiv] \lam274.20 
which is generally much stronger in active regions, but in the 
present spectrum \ion[Si vii] dominates, and provides $\simeq 57$\%
of the observed intensity (Table~\ref{lfunction.results}) if we
assume a density of $\log N_e = 9.15$. 

Two more lines are predicted to be bright enough to be observed,
and they are both emitted from the \orb[2s2p 5] \tm[1 P 1]
level. These two lines provide excellent density diagnostic ratios
when compared to the \tm[3 P ]--\tm[3 P ] lines discussed above. However,
one of them falls in the wavelength gap between the two EIS bands,
while the other is expected at 246.12~\AA\ and is lost under two
stronger blending lines of \ion[Si vi] and \ion[Fe xiii].

\subsection{\ion[Si viii]}
\label{sect.si8}

Two groups of \ion[Si viii] lines are expected in the EIS wavelength
ranges: the four $2s^22p^3$ $^2D_{J}$ -- $2s2p^4$ $^2D_{J^\prime}$
transitions between 276.8 and 277.1~\AA, and the two $2s^22p^3$
$^2P_{J}$ -- $2s2p^4$ $^2S_{1/2}$ transitions between 250 and
251~\AA. The latter two lines are very weak, but can be measured in
the present spectrum and will be discussed towards the end of this
section. The four $^2D$--$^2D$ transitions consist of a pair of 
strong transitions (3/2--3/2, 5/2--5/2 at 276.85 and 277.06~\AA,
respectively), and a pair of weak transitions (3/2--5/2, 5/2--3/2 at
276.87 and 277.04~\AA, respectively). These lines are blended with
lines of \ion[Mg vii] and \ion[Si vii] making it difficult to
extract the \ion[Si viii] line intensities, and we describe in detail
below the method used here.

\begin{figure}[h]
\epsscale{0.7}
\plotone{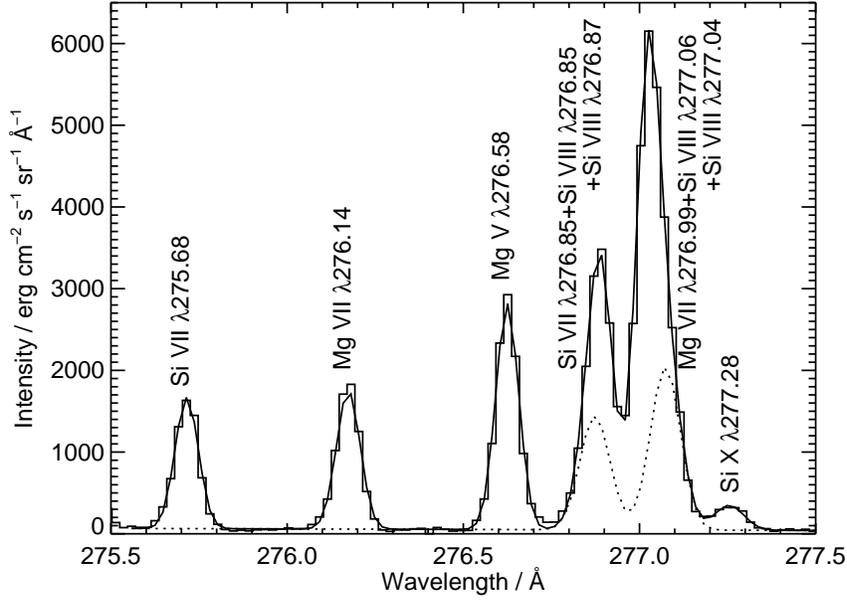}
\caption{The EIS spectrum from 275.5 to 277.5~\AA\ showing the
  emission lines that were simultaneously fit in order to derive line
  fit parameters for \ionx{Si}{viii}. The spectrum is shown as a
  histogram plot and the complete fit function is overlaid as a
  smooth, complete line. The \ionx{Si}{viii} components of the fit
  function are overlaid as a dotted line.}
\label{fig.si8}
\end{figure}

Fig.~\ref{fig.si8} shows the EIS spectrum in the vicinity of the
\ion[Si viii] $^2D$--$^2D$ transitions, with the different ion species
and lines indicated. Attempts to fit the lines simultaneously
with multiple Gaussians each with three free parameters (line peak,
centroid and width) fail due to the number of lines (7) between 276.8
and 277.3~\AA. The fitting process can be simplified significantly by
making use of the nearby \ion[Si vii] \lam275.68 and \ion[Mg vii]
\lam276.14 lines, which have fixed separations and ratios relative 
to the \ion[Si vii] and \ion[Mg vii] lines blending with the 
\ion[Si viii] lines.

The \ion[Si vii] \lam276.85/\lam275.68 branching ratio is 1.3. Also the 
separation of the two lines is accurately known from measurements of the 
$2s^22p^4$ $^3P_1$ -- $^3P_0$ transition at infrared wavelengths 
\citep{feucht97}. We thus allow the isolated \lam275.68 line to be 
freely fit, and then force \lam276.85  to have the same width (since 
the lines arise from the same ion), a peak 1.31 times that of \lam275.68, 
and a separation of 1.176~\AA. Similarly, \ion[Mg vii] \lam276.99/\lam276.14 
has a branching ratio of 2.99, and the wavelength separation is accurately 
known from infrared measurements of the $2s^22p^2$ $^3P_0$ -- $^3P_1$ 
wavelength \citep{kelly95}. The isolated \lam276.14 line is then used 
to determine the \lam276.99 parameters.

For \ion[Si viii], each of the four emission lines is forced to have
the same width, but this width is free to vary. The peaks and centroids 
of the two strong lines, \lam276.85 and \lam277.06, are free to vary, 
but those of the two weak lines, \lam276.87 and \lam277.04, are fixed 
relative to these lines. \lam276.85 and \lam277.04 share a common upper 
level and they have a branching ratio of 0.087 \citep[using the atomic 
data of][]{zhang99}, while their wavelength separation is accurately 
determined to be 0.193~\AA\ from the separation of the \ion[Si viii], 
\lam\lam1440, 1445 transitions in the far ultraviolet. Similarly 
\lam277.06 and \lam276.87 share a common upper level and have a branching 
ratio of 0.040, while their wavelength separation is also 0.193~\AA.

Two additional Gaussians are added to fit \ion[Mg v] \lam276.58 and
\ion[Si x] \lam277.26, each having completely free parameters. In
summary, the spectral region between 275.55 and 277.50~\AA\ is fit with
10 Gaussians together with a straight line for the background. There
are 21 free parameters in all, and the reduced $\chi^2$ value for the fit
is 2.3. The complete fit function is displayed in Fig.~\ref{fig.si8}
and Table~\ref{linelist} gives the line fit parameters for each of
the 10 emission lines. 

The wavelength separation of the two strong \ion[Si viii] lines is 
$0.201\pm 0.003$~\AA\ which is close to the value from \citet{edlen84} 
of 0.207~\AA, while Table~\ref{tbl.vel} shows that the velocities of 
the two lines are in good agreement with \ionx{Al}{viii}. Note that there
appears to be a jump in the spectrum redshift between the ions formed
at $\log\,T < 5.8$ and those formed at $\log\,T > 5.8$ of about 20~\kms.

The two $2s^22p^3$ $^2P_{J}$ -- $2s2p^4$ $^2S_{1/2}$ transitions have rest
wavelengths of 250.47 and 250.81~\AA\ and two weak lines are found close to 
these wavelengths in the spectrum. The velocity of \lam250.81 is consistent 
with \lam276.86 and \lam277.06 (Table~\ref{tbl.vel}), giving confidence in 
the identification. \lam250.47 is predicted to be a factor 0.65 of the
strength of \lam250.81, in good agreement with observations. However,
the line is very narrow and the velocity is significantly discrepant
with the other \ionx{Si}{viii} lines (Table~\ref{tbl.vel}). There is 
considerable density sensitivity in the \ion[Si viii] lines, and the
L-function method shows best agreement among all lines for $\log N_e=9.05\pm 0.30$.

\ion[Si viii]  emits several other strong spectral
lines in the 214--236~\AA\ wavelength range that are useful
for plasma diagnostics. Unfortunately, they all fall in the wavelength
gap between the two EIS bands and can not be used.

\subsection{\ion[Si ix]}
\label{sect.si9}

The only lines expected to be found in the EIS wavebands based on 
the CHIANTI atomic model are at 258.08 and 290.69~\AA, and both are
measured here. \lam258.08 is unblended but \lam290.69 lies in the
wing of a stronger \ion[Fe vii] line, and the combined feature was 
fit with two Gaussians forced to have the same width. These lines 
are strongly density sensitive relative to each other and the CHIANTI 
\ion[Si ix] model yields a value of $\log\,N_{\rm e}=9.20\pm 0.30$.

Other strong lines are predicted to fall in the \lam\lam220--230
wavelength range, that can not be observed with EIS.

\subsection{\ion[S viii]}
\label{sect.s8}

The \orb[2s 2]\orb[2p 5] \tm[2 P 3/2,1/2] -- \orb[2s2p 6] \tm[2 S 1/2]
doublet lines, \lam\lam198.55, 202.61, are found in the EIS short 
wavelength band and both are blended with \ionx{Fe}{xi} transitions. 
Thus, the results in Table~\ref{lfunction.results} can not be put in
an absolute scale. However, the derived DEM demonstrates that 
\ionx{S}{viii} provides the dominant contribution in both cases 
(Table~\ref{linelist}). \citet{brown08} also listed a \ionx{Fe}{xii} 
line as blending with \ionx{S}{viii} \lam198.55, but we find the 
predicted intensity for this line is negligible when using the DEM. 
If we use the reference wavelengths of \citet{robinson37} for
\ionx{S}{viii}, then the derived velocities are $+24.1$~\kms\ and
$+22.2$~\kms\ for \lam198.55 and \lam202.61, respectively, which are
consistent with the \ionx{Si}{viii} and \ionx{Al}{viii} velocities
(Table~\ref{tbl.vel}) and also confirm that \ionx{S}{viii} provides a
dominant contribution to both lines. Note that \ionx{S}{viii} has an
effective temperature of $\log\,T_{\rm eff}=5.83$, placing it between
\ionx{Si}{vii} and \ionx{Si}{viii}.

Since the \ionx{S}{viii} lines may be valuable for abundance studies
\citep[e.g.,][]{feldman09} we note that the \ionx{Fe}{xi} contributions
can more generally be estimated through branching ratios. Using the
atomic data from CHIANTI, \lam198.55/\lam189.14 has a theoretical ratio
of 0.80, while \lam202.63/\lam188.23 has a ratio of 0.016. \lam188.23
is the strongest \ionx{Fe}{xi} line observed by EIS, and \lam189.14
appears to be unblended.





\subsection{\ion[Cr vii]}
\label{sect.cr7}

\ion[Cr vii] lines have been identified in laboratory spectra
\citep{gabriel66,ekberg76}, but never previously seen in solar
spectra. The ion is isoelectronic with \ion[Fe ix] and the analogous
transition to the strong \lam171.07 line of \ion[Fe ix] is found at
202.83~\AA, which matches a strong line at this
wavelength. An image formed from the line looks similar to
\ion[Fe vii] and \ion[Fe viii] images, which is consistent with the
predicted temperature of maximum ionization, $\log\,T_{\rm max}=5.7$,
of \ion[Cr vii]. In addition, using the laboratory
wavelength of \citet{ekberg76} yields a velocity of $+42.9$~\kms, in
good agreement with \ionx{Si}{vii} which is formed at a similar
temperature (Table~\ref{tbl.vel}). \citet{brown08} identify the
\ionx{O}{iv} \lam202.885 line in their spectra which is close in
wavelength to the \ionx{Cr}{vii} line, however the CHIANTI model for
\ionx{O}{iv} predicts this transition to be around a factor 40 weaker
than \ionx{O}{iv} \lam279.933 and so in the present spectrum it can
safely be ignored. \citet{landi09} have computed new atomic data for
\ion[Cr vii] and the emission measure distribution yields a line 
intensity in reasonable agreement with the observations considering 
the uncertainties in the atomic calculations. We are thus confident 
that the 202.83~\AA\ emission line is due to \ion[Cr vii].

By comparing with the \ion[Fe ix] atomic model, we expect also to
find additional lines at 258.65, 259.18, 261.31 and 288.90~\AA\ that
will be around 5--20~\%\ of the \lam202.83 line. There are indeed
lines at these wavelengths except at \lam288.90; Table~\ref{lfunction.results} 
shows that all these weaker lines are brighter than predicted. Two of
these lines are predicted to be strongly density sensitive but no
common crossing point can be identified: Table~\ref{lfunction.results} 
results have been obtained assuming that $\log N_e=9.15$. No clear
candidates for blending are available for any of the lines except 
\lam259.18,, which is blended by an \ion[Al vii] line (Section~\ref{sect.al7}). 
Also, they are too weak to provide an intensity map that can be matched
with a temperature class, so the identification of these lines
remains to be confirmed.

\subsection{\ion[Cr viii]}
\label{sect.cr8}

With the identification of \ion[Cr vii] \lam202.83, we also expect to
find \ion[Cr viii] \lam205.01 which is the analogous transition to the 
strong \lam174.53 line of \ion[Fe x]. A line is found at 205.05~\AA\ 
which is around a factor 2 weaker than \lam202.83 and, based on image 
inspection, is formed close to \ion[Fe vii] and \ion[Fe viii]. We thus 
identify this with \ion[Cr viii]. Converting the measured wavelength 
to a velocity gives $+62.9$~\kms, significantly different to 
\ionx{Si}{viii} and \ionx{Al}{viii} which are formed at a similar 
temperature (Table~\ref{tbl.vel}). However, the reference wavelengths 
for \ionx{Cr}{viii} come from the laboratory spectra obtained by
\citet{fawcett66}  but are only accurate to $\pm$0.05~\AA\ 
\citep[see also][]{gabriel66}. 

Other \ion[Cr viii] lines are found in the EIS range, and the comparison
with  \lam205.01 is reported in Table~\ref{lfunction.results}. Good
agreement is found; the lines are also density sensitive relative to 
each other, and the best agreement is found for $\log N_e=9.45$, a bit
higher than the values found for other ions. The \lam211.48 is also
reported as \ion[Ni xi]: while \ion[Ni xi] is predicted to provide $\simeq
50$\% of the total intensity, \ion[Cr viii] accounts for all of it. The
reason for this discrepancy is not clear, but it is likely to be found 
in the atomic data.

\subsection{\ion[Mn viii]}
\label{sect.mn8}

\ion[Mn viii] is isoelectronic with \ion[Fe ix], and its spectrum is also
dominated by the strong singlet transition \orb[3s 2]\orb[3p 6] \tm[1 S ] -
\orb[3s 2]\orb[3p 5]\orb[3d ] \tm[1 P ], which is identified as \lam185.462.
Using the new atomic data from \citet{landi09}, the predicted intensity of 
this line agrees well with observations, however the velocity derived using 
the reference wavelength of \citet{smitt83} is discrepant with other ions 
of the same temperature such as \ionx{Si}{vii} and \ionx{Mg}{vii} by around 
20~\kms. The velocity instead is more consistent with the hotter ions 
\ionx{Si}{viii} and \ionx{Al}{viii}, perhaps suggesting the ion balance 
for \ion[Mn viii] is yielding too low a temperature.

Other, weaker \ion[Mn viii] lines are predicted to fall in the EIS wavelength 
ranges, the strongest of which is identified at 263.200~\AA\ with a weak line 
whose intensity map is consistent with a cool line. Table~\ref{lfunction.results} 
reports the comparison between the two lines. They are strongly density sensitive
for $\log Ne > 9.5$, but at lower densities their ratio is more constant; \lam263.20 
is brighter than predicted by the comparison with the strong singlet line, so that 
some unidentified blend has to provide almost 60\% of the total observed intensity.

\subsection{\ion[Mn ix]}
\label{sect.mn9}

Using the atomic data of \citet{landi09}, the DEM yields  three \ion[Mn ix]
lines that are potentially observable in the present spectrum, and wavelength 
matches are found for each. \lam188.48 is the strongest line and is the 
analogous transition to \ionx{Fe}{x} \lam174.53. It lies in a crowded part 
of the spectrum and it could provide a contribution to either \ionx{Fe}{vii} 
\lam188.40 or \ionx{Fe}{ix} \lam188.50 (observed at 188.425 and 188.507~\AA, 
respectively). The reference wavelength \citep[due to][]{fawcett66} is only 
accurate to $\pm 0.05$~\AA\ and so it is not possible to clearly identify 
\ionx{Mn}{ix} with either of the observed lines. We note, however, that 
\citet{young09a} find that \ionx{Fe}{vii} does not fully account for the 
strength of the observed line at 188.425~\AA\ and so in Table~\ref{linelist} 
we identify \ionx{Mn}{ix} with this line.

The two other potential \ionx{Mn}{ix} identifications are \lam191.60
and \lam199.32 which are the analogous transitions to \ionx{Fe}{x}
\lam177.24 and \lam184.54, respectively. Using the \citet{fawcett66}
reference wavelengths yields velocities of $+18.8$ and $+7.5$~\kms,
respectively, which give confidence in the identifications although 
we note again the low precision of the laboratory wavelengths. The
intensities predicted from the DEM are significantly below the 
observed intensities, however images formed in both the lines are 
consistent with the expected formation temperature of \ionx{Mn}{ix} 
($\log\,T_{\rm eff}=5.86$). Given the uncertainties in the identifications 
and blends, the results in Table~\ref{lfunction.results} can not be put 
on an absolute scale.

New, more precise, laboratory wavelength measurements would be
valuable for confirming the \ionx{Mn}{ix} identifications.

\subsection{\ion[Ni xi]}
\label{sect.ni11}

\ion[Ni xi] has the same atomic structure as \ion[Fe ix] and its spectrum
is dominated by the strong singlet line, analogous to \ion[Fe ix] \lam171.07,
which lies outside the EIS wavelength range. Three \ion[Ni xi] lines are 
identified in this spectrum, one of which for the first time as blending 
an \ion[Al v] transition at \lam278.73. \ion[Ni xi] is expected to provide
$\simeq 13$\% of the total intensity, bringing the combined predicted intensity 
of the blended feature much closer to observations. Due to the \ion[Al v]
blend, the identification of this line does not allow us to determine an
accurate wavelength for this transition and calculate from it the energy
of the upper level.

The other two lines are identified near the edge of the EIS short wavelength
band. These two lines are analogous to \ion[Fe ix] \lam\lam241.74, 244.91,
which have been observed in the past and used for density diagnostics due
to the strong density sensitivity of their intensity ratio. In the present
spectrum, \lam211.48 is blended with a \ion[Cr viii] transition. Some problem
is found here, as the \ion[Cr viii] line accounts for all the observed intensity,
while Table~\ref{lfunction.results} predicts \ion[Ni xi] to provide $\simeq 50$\% 
of the total intensity.

\section{Conclusions}
\label{sect.conclusions}

In the present work we have analyzed a full EIS spectral scan of a portion
of an active region where the plasma emission was enhanced at transition
region temperatures. We first measured the DEM and electron density of the
plasma, and used the results to develop a complete atlas of the emitting
spectrum, and to compare observed line intensities with predicted values
from the CHIANTI database. The \ion[Fe vii-ix] lines measured in the
present spectrum have been analyzed in a separate paper \citep{young09a}.

While most of the lines identified in the spectrum were observed in other
occasions by EIS, the strong enhancement of the emission at temperatures
in the $\log T=5.5$--5.9 range has allowed us to identify several lines
never observed in solar spectra. These lines, sometimes identified in 
laboratory spectra, are usually too faint to be detected but in special
plasmas like the one we studied can provide valuable diagnostic tools to
measure the physical properties of the emitting plasma.

The observed spectrum was also used to carry out a systematic assessment 
of the accuracy of CHIANTI emissivities, as well as of the diagnostic 
application, for transition region ions. The brightness of the transition
region emission has allowed us to carry out such a comparison including
more lines and ions and with better accuracy than possible with standard
active region or quiet Sun spectra.

We find that CHIANTI emissivities are almost always in excellent agreement
with observations. We identified blends for several lines, and discussed
the diagnostic application of many of the lines reported in the present
atlas.

\acknowledgements

The work of EL is supported by the NNG06EA14I, NNH06CD24C and other NASA grants.

\begin{deluxetable}{rrrrrrlllllcr}
\tabletypesize{\tiny}
\tablecaption{Line list. $\lambda$ and $\sigma_\lambda$ are the line wavelength and its uncertainty 
(in \AA\ and m\AA, respectively); $I$ and $\sigma_I$ are the line intensity and its uncertainty
(in erg~cm$^{-2}$s$^{-1}$sr$^{-1}$; $W$ and $\sigma_W$ are the line width and its uncertainty (in m\AA).
For some strong lines, $\sigma_\lambda<0.5$~m\AA\ and so the rounded $\sigma_\lambda$ given
in the table is 0. \label{linelist}}
\tablewidth{0pt}
\tablehead{
  \colhead{$\lambda$} & \colhead{$\sigma_\lambda$} & \colhead{$I$} & \colhead{$\sigma_I$}& \colhead{$W$} & \colhead{$\sigma_W$} & \colhead{Class} & \colhead{Ion} & \colhead{$\lambda_{\rm ref}$} & \colhead{Ref.} & \colhead{Transition} & \colhead{$\log\,T_{\rm max}$} & \colhead{$I_{\rm pred}$} 
}
\startdata
171.070 & 1 & 6779.7 & 356.2 & 83.7 & 1.1 & E & \ion[Fe ix] & 171.073 & 1 & $3s^23p^6$ $^1S_{0}$ -- $3s^23p^53d$ $^1P_{1}$ &    5.7   &  9070.0  \\ 
172.926 & 15 & 199.5 & 108.4 & 63.1 & 34.3 &  & \ion[O vi] & 172.936 & 1 & $2p$ $^2P_{1/2}^{\rm o}$ -- $3d$ $^2D_{3/2}$ &  5.6  &  105.0 \\ 
173.120 & 17 & 300.5 & 127.4 & 99.3 & 42.1 &  & \ion[O vi] & 173.080 & 1 & $2p$ $^2P_{3/2}^{\rm o}$ -- $3d$ $^2D_{5/2}$ &  5.6  &  190.0 \\ 
 &  &  &  &  &  &  & \ion[O vi] & 173.095 & 1 & $2p$ $^2P_{3/2}^{\rm o}$ -- $3d$ $^2D_{3/2}$ & 5.6  & 21.0  \\ 
174.523 & 1 & 1810.5 & 87.9 & 88.6 & 1.1 & F & \ion[Fe x] & 174.531 & 1 & $3s^23p^5$ $^2P_{3/2}^{\rm o}$ -- $3s^23p^4(^3P)3d$ $^2D_{5/2}$ &    6.0   &  1750.0  \\ 
175.265 & 1 & 232.8 & 12.4 & 76.7 & 1.1 &  & \ion[Fe x] & 175.263 & 1 & $3s^23p^5$ $^2P_{1/2}^{\rm o}$ -- $3s^23p^4(^3P)3d$ $^2D_{3/2}$ &    6.0   &  260.0  \\ 
176.762 & 4 & 191.4 & 32.7 & 57.9 & 9.5 &  & \ion[Fe vii] & 176.744 & 3 & $3p^63d^2$ $^3F_4$ -- $3p^53d^3(^4F)$ $^3F_4^{\rm o}$ &  5.6  &   108.0 \\ 
176.968 & 4 & 446.7 & 42.8 & 109.5 & 10.1 &  & \ion[Fe ix] & 176.959 & 2 & $3s^23p^53d$ $^3F_{4}^{\rm o}$ -- $3s^23p^4(^1D)3d^2$ $^3D_{3}$ &  5.7  &   274.0 \\ 
 &  &  &  &  &  &   & \ion[Fe vii] & 176.928 & 3 & $3p^63d^2$ $^3F_3$ -- $3p^53d^3(^4F)$ $^3F_3^{\rm o}$ &  5.6  &    74.5 \\ 
177.236 & 2 & 1101.9 & 52.8 & 80.3 & 3.3 & F & \ion[Fe x] & 177.240 & 1 & $3s^23p^5$ $^2P_{3/2}^{\rm o}$ -- $3s^23p^4(^3P)3d$ $^2P_{3/2}$ &    6.0   &  957.0  \\ 
 &  &  &  &  &  &  & \ion[Fe vii] & 177.172 & 3 & $3p^63d^2$ $^3F_2$ -- $3p^53d^3(^4F)$ $^3F_2^{\rm o}$ &  5.6  &    52.9 \\ 
177.603 & 6 & 148.2 & 26.1 & 86.9 & 15.3 &  & \ion[Fe ix] & 177.594 & 2 & $3s^23p^53d$ $^3F_{3}^{\rm o}$ -- $3s^23p^4(^1D)3d^2$ $^3D_{2}$ &  5.7  &   139.0  \\ 
178.708 & 2 & 68.5 & 5.0 & 84.5 & 4.1 \\ 
178.994 & 3 & 76.0 & 5.2 & 115.5 & 5.5 \\ 
179.245 & 2 & 45.8 & 4.0 & 60.3 & 3.6 \\ 
179.740 & 3 & 56.3 & 4.7 & 108.5 & 6.2 & G & \ion[Fe xi] & 179.764 & 1 & $3s^23p^4$ $^1D_{2}$ -- $3s^23p^3(^2D)3d$ $^1F_{3}$ &    6.0   &  35.2  \\ 
180.402 & 1 & 840.8 & 43.8 & 95.9 & 1.9 & G & \ion[Fe xi] & 180.408 & 1 & $3s^23p^4$ $^3P_{2}$ -- $3s^23p^3(^4S)3d$ $^3D_{3}$ &    6.0   &  739.0  \\ 
 &  &  &  &  &  &  & \ion[Fe x] & 180.441 & 1 & $3s^23p^5$ $^2P_{1/2}^{\rm o}$ -- $3s^23p^4(^3P)3d$ $^2P_{1/2}$ &    6.0   &  148.0  \\ 
180.618 & 4 & 34.6 & 3.8 & 74.7 & 6.9 & G & \ion[Fe xi] & 180.600 & 1 & $3s^23p^4$ $^3P_{1}$ -- $3s^23p^3(^4S)3d$ $^3D_{1}$ &    6.0   &  42.5  \\ 
181.119 & 3 & 34.1 & 3.9 & 74.7 & 6.8 & G & \ion[Fe xi] & 181.137 & 1 & $3s^23p^4$ $^3P_{0}$ -- $3s^23p^3(^4S)3d$ $^3D_{1}$ &    6.0   &  59.7  \\ 
181.374 & 5 & 16.8 & 2.8 & 66.2 & 9.9 \\ 
181.622 & 12 & 12.4 & 5.2 & 67.4 & 28.1 \\ 
181.744 & 6 & 22.2 & 5.1 & 65.3 & 15.0 \\ 
182.099 & 9 & 23.1 & 8.8 & 56.0 & 0.0 &  & \ion[Fe vii] & 182.071 & 3 & $3p^63d^2$ $^3F_4$ -- $3p^53d^3(^2F)$ $^3D_3^{\rm o}$ &  5.6  &     4.7 \\ 
182.171 & 3 & 151.3 & 13.7 & 79.2 & 7.2 & G & \ion[Fe xi] & 182.169 & 1 & $3s^23p^4$ $^3P_1$ -- $3s^23p^3(^4S)3d$ $^3D_2^{\rm o}$ &  6.0  &  149.0  \\ 
182.308 & 3 & 58.2 & 7.2 & 75.3 & 9.2 & F & \ion[Fe x] & 182.307 & 1 & $3s^23p^5$ $^2P_{1/2}^{\rm o}$ -- $3s^23p^4(^3P)3d$ $^2P_{3/2}$ &    6.0   &  25.8  \\ 
182.430 & 7 & 21.6 & 5.5 & 68.3 & 17.2 \\ 
182.755 & 7 & 10.7 & 3.6 & 54.5 & 18.5 \\ 
182.946 & 7 & 26.2 & 4.7 & 93.0 & 16.9 \\ 
183.163 & 8 & 11.6 & 4.3 & 60.6 & 22.3 \\ 
183.345 & 6 & 13.0 & 3.1 & 57.3 & 13.5 \\ 
183.467 & 7 & 11.5 & 3.2 & 62.5 & 17.5 \\ 
183.566 & 5 & 12.0 & 2.8 & 47.9 & 11.2 &  & \ion[Fe vii] & 183.539 & 3 & $3p^63d^2$ $^3P_1$ -- $3p^53d^3(^4P)$ $^3P_2^{\rm o}$ &  5.6  &    8.9  \\
183.849 & 2 & 90.9 & 5.4 & 65.5 & 3.9 & C & \ion[Fe vii] & 183.825 & 3 & $3p^63d^2$ $^3P_2$ -- $3p^53d^3(^4P)$ $^3P_2^{\rm o}$ &  5.6  &    33.0 \\
183.951 & 2 & 96.1 & 5.4 & 68.8 & 3.9 & A & \ion[O vi] & 183.939 & 9 & $1s^22p$ $^2P_{1/2}^{\rm o}$ -- $1s^23s$ $^2S_{1/2}$ &  5.6 &  65.8 \\ 
184.141 & 1 & 145.3 & 5.9 & 74.2 & 2.5 & A & \ion[O vi] & 184.119 & 9 & $1s^22p$ $^2P_{3/2}^{\rm o}$ -- $1s^23s$ $^2S_{1/2}$ &  5.6 &  132.0 \\ 
184.426 & 4 & 36.4 & 4.6 & 70.0 & 8.9 & F & \ion[Fe xi] & 184.412 & 1 & $3s^23p^4$ $^1S_0$ -- $3s^23p^3(^2P)3d$ $^1P_1^{\rm o}$ &  &  1.4 \\ 
184.542 & 1 & 504.3 & 9.2 & 76.7 & 1.2 & F & \ion[Fe x] & 184.537 & 1 & $3s^23p^5$ $^2P_{3/2}^{\rm o}$ -- $3s^23p^4(^1D)3d$ $^2S_{1/2}$ &    6.0   &  397.0  \\ 
184.777 & 3 & 36.3 & 3.1 & 70.0 & 6.0 &  & \ion[Ne v] & 184.735 & 1 & $2s^22p^2$ $^1S_{0}$ -- $2s^22p3s$ $^1P_{1}$ &    5.6   &  7.7  \\ 
 &  &  &  &  &  &  & \ion[Fe xi] & 184.803 & 1 & $3s^23p^4$ $^1D_{2}$ -- $3s^23p^3(^2D)3d$ $^1D_{2}$ &    6.0   &  21.3  \\ 
 &  &  &  &  &  &  & \ion[Fe vii] & 184.752 & 3 & $3p^63d^2$ $^3P_0$ -- $3p^53d^3(^4P)$ $^3P_1^{\rm o}$ &  5.6  &    8.3 \\ 
184.922 & 3 & 20.2 & 2.8 & 61.9 & 8.6 &  & \ion[Ne vi] & 184.945 & 1 & $2s2p^2$ $^2S_{1/2}$ -- $2s^23p$ $^2P_{3/2}^{\rm o}$ &    5.7   &  8.7  \\ 
 &  &  &  &  &  &  & \ion[Fe vii] & 184.886 & 3 & $3p^63d^2$ $^3P_1$ -- $3p^53d^3(^4P)$ $^3P_1^{\rm o}$ &  5.6  &     7.8 \\ 
185.232 & 0 & 1503.2 & 13.8 & 67.8 & 0.4 & D & \ion[Fe viii] & 185.213 & 14 & $3p^63d$ $^2D_{5/2}$ -- $3p^53d^2(^3F)$ $^2F_{7/2}^{\rm o}$ &    5.7   &  2760.0  \\ 
185.467 & 2 & 116.9 & 5.3 & 87.0 & 3.9 & C--D & \ion[Mn viii] & 185.455 & 12 & $3s^2 3p^6$ $^1S_{0}$ -- $3s^2 3p^5 3d$ $^1P_{1}^{\rm o}$ &    5.7   &  105.0  \\ 
185.574 & 2 & 66.1 & 3.9 & 59.6 & 3.6 & B--C & \ion[Fe vii] & 185.547 & 3 & $3p^63d^2$ $^1D_2$ -- $3p^53d^3(^2G)$ $^1F_3^{\rm o}$ &  5.6  &    36.0 \\ 
185.780 & 10 & 7.7 & 2.1 & 79.9 & 9.0 &  & \ion[O v] & 185.745 & 1 & $2s2p$ $^1P_{1}^{\rm o}$ -- $2p3p$ $^1D_{2}^{\rm o}$ &    5.6   &  3.3  \\ 
186.004 & 5 & 18.5 & 2.6 & 79.9 & 9.0 & C--E \\ 
186.142 & 7 & 12.4 & 2.2 & 79.9 & 9.0 \\ 
186.624 & 1 & 1139.9 & 14.0 & 67.4 & 0.8 & D & \ion[Fe viii] & 186.601 & 14 & $3p^63d$ $^2D_{3/2}$ -- $3p^53d^2(^3F)$ $^2F_{5/2}^{\rm o}$ &    5.7   &  1890.0  \\ 
186.692 & 6 & 43.8 & 7.2 & 53.7 & 8.2 &  & \ion[Fe vii] & 186.657 & 3 & $3p^63d^2$ $^1D_2$ -- $3p^53d^3({\rm b}~^2D)$ $^1D_2^{\rm o}$ &  5.6  &    34.4 \\ 
186.877 & 1 & 174.2 & 4.3 & 93.1 & 2.3 & H & \ion[Fe xii] & 186.854 & 1 & $3s^23p^3$ $^2D_{3/2}^{\rm o}$ -- $3s^23p^2(^3P)3d$ $^2F_{5/2}$ &    6.1   &  26.5  \\ 
 &  &  &  &  &  &  & \ion[Fe xii] & 186.887 & 1 & $3s^23p^3$ $^2D_{5/2}^{\rm o}$ -- $3s^23p^2(^3P)3d$ $^2F_{7/2}$ &    6.1   &  95.8  \\ 
 &  &  &  &  &  &  & \ion[Fe vii] & 186.868 & 3 & $3p^63d^2$ $^3P_2$ -- $3p^53d^3(^2G)$ $^1F_3^{\rm o}$ &  5.6  &    7.9 \\ 
187.181 & 35 & 16.1 & 10.6 & 103.7 & 68.6 \\ 
187.264 & 3 & 114.0 & 10.5 & 66.9 & 4.4 & C & \ion[Fe viii] & 187.237 & 14 & $3p^63d$ $^2D_{5/2}$ -- $3p^53d^2(^3F)$ $^2F_{5/2}^{\rm o}$ &    5.7   &  86.4  \\ 
 &  &  &  &  &  &  & \ion[Fe vii] & 187.235 & 3 & $3p^63d^2$ $^1D_2$ -- $3p^53d^3(^2F)$ $^3D_3^{\rm o}$ &  5.6  &    9.0 \\ 
187.433 & 7 & 11.6 & 3.1 & 81.5 & 21.5 \\ 
187.562 & 8 & 9.0 & 2.5 & 75.2 & 21.0 \\ 
187.714 & 2 & 22.3 & 2.6 & 63.8 & 7.5 & B & \ion[Fe vii] & 187.692 & 3 & $3p^63d^2$ $^3P_1$ -- $3p^53d^3({\rm b}~^2D)$ $^1D_2^{\rm o}$ &  5.6  &     4.4 \\ 
187.848 & 5 & 7.1 & 1.8 & 58.2 & 15.2 \\ 
187.971 & 1 & 89.8 & 3.3 & 89.8 & 3.1 & E  \\ 
188.144 & 10 & 42.5 & 10.5 & 95.0 & 0.0 \\ 
188.216 & 1 & 386.6 & 18.2 & 78.6 & 3.7 & G & \ion[Fe xi] & 188.232 & 1 & $3s^23p^4$ $^3P_{2}$ -- $3s^23p^3(^2D)3d$ $^3P_{2}^{\rm o}$ &  6.0  &  361.0  \\ 
188.303 & 2 & 257.0 & 12.2 & 77.0 & 3.6 & G & \ion[Fe xi] & 188.299 & 1 & $3s^23p^4$ $^3P_{2}$ -- $3s^23p^3(^2D)3d$ $^1P_{1}$ &    6.0   &  131.0  \\ 
188.424 & 3 & 71.8 & 7.6 & 66.7 & 7.1 & C & \ion[Mn ix] & 188.480 & 1 & $3s^23p^5$ $^2P_{3/2}^{\rm o}$ -- $3s^23p^4(^3P)3d$ $^2D_{5/2}$ &    6.0   &  21.6  \\ 
 &  &  &  &  &  &  & \ion[Fe vii] & 188.396 & 3 & $3p^63d^2$ $^1D_2$ -- $3p^53d^3(^2F)$ $^3D_2^{\rm o}$ &  5.6  &    12.5 \\ 
188.507 & 1 & 375.5 & 13.2 & 70.2 & 2.5 & E & \ion[Fe ix] & 188.497 & 16 & $3s^23p^53d$ $^3F_{4}$ -- $3s^23p^4(^3P)3d^2$ $^3G_{5}$ &    5.7   &  396.0  \\ 
188.603 & 3 & 72.6 & 10.7 & 69.1 & 10.2 & C & \ion[Fe vii] & 188.576 & 3 & $3p^63d^2$ $^3P_2$ -- $3p^53d^3(^2F)$ $^3D_3^{\rm o}$ &  5.6  &    36.6 \\ 
188.685 & 7 & 31.7 & 8.0 & 72.8 & 18.3 & E,H & \ion[Fe ix] & 188.686 & 1 & $3s^23p^53d$ $^3F_{4}$ -- $3s^23p^4(^3P)3d^2$ $^3G_{4}$ &    5.7   &  22.2  \\ 
188.768 & 15 & 7.5 & 2.7 & 56.0 & 0.0 \\ 
188.833 & 4 & 42.9 & 3.8 & 75.2 & 6.7 \\ 
189.001 & 3 & 22.0 & 2.4 & 86.1 & 9.3 \\ 
189.127 & 21 & 17.7 & 7.0 & 74.4 & 29.4 & G & \ion[Fe xi] & 189.130 & 1 & $3s^23p^4$ $^3P_{1}$ -- $3s^23p^3(^2D)3d$ $^3P_{1}^{\rm o}$ &  6.0  &  25.0  \\ 
189.359 & 1 & 30.5 & 1.7 & 60.6 & 3.4 \\ 
189.481 & 1 & 53.1 & 2.1 & 66.7 & 2.7 & B & \ion[Fe vii] & 189.450 & 3 & $3p^63d^2$ $^3P_1$ -- $3p^53d^3(^2F)$ $^3D_2^{\rm o}$ &  5.6  &    19.5 \\ 
189.596 & 4 & 8.0 & 1.6 & 51.6 & 10.1 & E & \ion[Fe ix] & 189.582 & 1 & $3s^23p^53d$ $^3F_{3}^{\rm o}$ -- $3s^23p^4(^3P)3d^2$ $^3G_{3}$ &  5.7   &  26.0  \\ 
189.715 & 5 & 6.3 & 1.5 & 52.7 & 13.0 & G & \ion[Fe xi] & 189.719 & 1 & $3s^23p^4$ $^3P_{0}$ -- $3s^23p^3(^2D)3d$ $^3P_{1}^{\rm o}$ &  6.0  &   19.2  \\ 
189.952 & 1 & 236.7 & 4.8 & 74.8 & 1.5 & E & \ion[Fe ix] & 189.941 & 16 & $3s^23p^53d$ $^3F_{3}$ -- $3s^23p^4(^3P)3d^2$ $^3G_{4}$ &    5.7   &  241.0  \\ 
190.046 & 1 & 181.9 & 4.4 & 72.0 & 1.8 & F & \ion[Fe x] & 190.037 & 1 & $3s^23p^5$ $^2P_{1/2}^{\rm o}$ -- $3s^23p^4(^1D)3d$ $^2S_{1/2}$ &    6.0   &  112.0  \\ 
190.164 & 5 & 3.4 & 1.0 & 43.9 & 13.3 \\ 
190.359 & 4 & 18.2 & 2.1 & 74.4 & 1.9 \\ 
190.909 & 5 & 19.2 & 2.4 & 76.6 & 1.6 \\ 
191.041 & 11 & 7.1 & 1.9 & 76.6 & 1.6 & H & \ion[Fe xii] & 191.049 & 1 & $3s^23p^3$ $^2P_{3/2}^{\rm o}$ -- $3s^23p^2(^3P)3d$ $^2D_{5/2}$ &    6.1   &  5.3  \\ 
191.227 & 2 & 105.2 & 6.7 & 76.6 & 1.6 & E,H & \ion[Fe ix] & 191.216 & 16 & $3s^23p^53d$ $^3F_{2}$ -- $3s^23p^4(^3P)3d^2$ $^3G_{3}$ &    5.7   &  105.0  \\ 
191.411 & 7 & 11.5 & 2.1 & 76.6 & 1.6 \\ 
191.612 & 4 & 37.7 & 3.5 & 76.6 & 1.6 & D & \ion[Mn ix] & 191.630 & 1 & $3s^23p^5$ $^2P_{3/2}^{\rm o}$ -- $3s^23p^4(^3P)3d$ $^2P_{3/2}$ &    6.0   &  10.4  \\ 
191.707 & 13 & 8.5 & 2.2 & 76.6 & 1.6 \\ 
191.814 & 10 & 8.9 & 2.0 & 76.6 & 1.6 \\ 
192.026 & 3 & 86.5 & 6.8 & 76.6 & 1.6 & D,H & \ion[Fe viii] & 192.004 & 14 & $3p^63d$ $^2D_{3/2}$ -- $3p^53d^2(^1S)$ $^2P_{1/2}^{\rm o}$ &    5.7   &  42.5  \\ 
 &  &  &  &  &  &  & \ion[Fe vii] & 192.006 & 3 & $3p^63d^2$ $^1D_2$ -- $3p^53d^3({\rm a}~^2D)$ $^1D_2^{\rm o}$ &  5.6  &  3.0  \\ 
192.114 & 4 & 60.4 & 5.8 & 76.6 & 1.6 \\ 
192.313 & 7 & 32.2 & 5.9 & 76.6 & 1.6 \\ 
192.386 & 3 & 100.5 & 8.2 & 76.6 & 1.6 & H & \ion[Fe xii] & 192.394 & 1 & $3s^23p^3$ $^4S_{3/2}^{\rm o}$ -- $3s^23p^2(^3P)3d$ $^4P_{1/2}$ &    6.1   &  135.0  \\ 
192.642 & 1 & 76.7 & 2.0 & 74.2 & 1.9 & E--F  \\ 
192.777 & 1 & 14.4 & 0.3 & 65.5 & 1.4 &  & \ion[O v] & 192.750 & 1 & $2s2p$ $^3P_0^{\rm o}$ -- $2s3d$ $^3D_1$ &  5.6  &  17.0 \\
192.808 & 1 & 101.3 & 2.7 & 79.8 & 2.1 & G & \ion[Fe xi] & 192.830 & 1 & $3s^23p^4$ $^3P_{1}$ -- $3s^23p^3(^2D)3d$ $^3P_{2}$ &    6.1   &  75.3  \\
192.824 & 1 & 27.0 & 0.6 & 65.5 & 1.4 &  & \ion[O v] & 192.797 & 1 & $2s2p$ $^3P_1^{\rm o}$ -- $2s3d$ $^3D_2$ &  5.6  &  30.7 \\
192.828 & 1 & 10.8 & 0.2 & 65.5 & 1.4 &  & \ion[O v] & 192.801 & 1 & $2s2p$ $^3P_1^{\rm o}$ -- $2s3d$ $^3D_1$ &  5.6  &  12.7 \\
192.931 & 1 & 82.2 & 1.9 & 65.5 & 1.4 & A & \ion[O v] & 192.904 & 1 & $2s2p$ $^3P_{2}^{\rm o}$ -- $2s3d$ $^3D_{3}$ &    5.6   &  99.7  \\
192.938 & 1 & 9.0 & 0.2 & 65.5 & 1.4 &  & \ion[O v] & 192.911 & 1 & $2s2p$ $^3P_2^{\rm o}$ -- $2s3d$ $^3D_2$ &  5.6  &  10.2 \\ 
193.721 & 2 & 71.2 & 5.6 & 94.4 & 5.0 & F & \ion[Fe x] & 193.715 & 1 & $3s^23p^5$ $^2P_{3/2}^{\rm o}$ -- $3s^23p^4(^1S)3d$ $^2D_{5/2}$ &    6.0   &  23.4  \\ 
193.988 & 2 & 44.7 & 4.3 & 60.7 & 3.9 & D & \ion[Fe viii] & 193.967 & 14 & $3p^63d$ $^2D_{3/2}$ -- $3p^64p$ $^2P_{3/2}^{\rm o}$ &    5.7   &  53.2  \\ 
194.319 & 5 & 12.8 & 2.2 & 72.5 & 12.2 \\ 
194.680 & 0 & 457.9 & 4.1 & 65.9 & 0.4 & D & \ion[Fe viii] & 194.662 & 14 & $3p^63d$ $^2D_{5/2}$ -- $3p^64p$ $^2P_{3/2}^{\rm o}$ &    5.7   &  524.0  \\ 
194.816 & 1 & 109.0 & 2.2 & 91.0 & 1.9 & C--E  \\ 
195.115 & 1 & 403.6 & 23.2 & 96.1 & 2.6 & H & \ion[Fe xii] & 195.119 & 1 & $3s^23p^3$ $^4S_{3/2}^{\rm o}$ -- $3s^23p^2(^3P)3d$ $^4P_{5/2}$ &    6.1   &  421.0  \\ 
 &  &  &  &  &  &  & \ion[Fe xii] & 195.179 & 1 & $3s^23p^3$ $^2D_{3/2}^{\rm o}$ -- $3s^23p^2(^1D)3d$ $^2D_{3/2}$ &    6.1   &  12.0  \\ 
195.415 & 0 & 224.8 & 3.0 & 64.0 & 0.8 & C & \ion[Fe vii] & 195.391 & 3 & $3p^63d^2$ $^3F_4$ -- $3p^53d^3(^2H)$ $^3G_5^{\rm o}$ &  5.6  &    70.7 \\ 
195.506 & 1 & 116.8 & 2.5 & 63.2 & 1.4 & C & \ion[Fe vii] & 195.485 & 2 & $3p^63d^2$ $^3F_3$ -- $3p^53d^3(^2H)$ $^3G_4^{\rm o}$ &  5.6  &    47.8 \\ 
195.753 & 2 & 20.2 & 1.3 & 61.9 & 3.9 & E, G--H \\ 
195.993 & 0 & 312.3 & 3.8 & 66.4 & 0.8 & D & \ion[Fe viii] & 195.972 & 14 & $3p^63d$ $^2D_{3/2}$ -- $3p^64p$ $^2P_{1/2}^{\rm o}$ &    5.7   &  357.0  \\ 
196.074 & 1 & 80.4 & 2.8 & 60.5 & 2.1 & B--C & \ion[Fe vii] & 196.046 & 3 & $3p^63d^2$ $^3F_2$ -- $3p^53d^3(^2H)$ $^3G_3^{\rm o}$ &  5.6  &    24.5 \\ 
196.239 & 0 & 113.9 & 2.0 & 57.9 & 0.9 & B--C & \ion[Fe vii] & 196.217 & 2 & $3p^63d^2$ $^1G_4$ -- $3p^53d^3(^2H)$ $^1H_5^{\rm o}$ &  5.6  &    38.6 \\ 
196.458 & 1 & 35.8 & 1.4 & 62.5 & 4.0 & B,L & \ion[Fe vii] & 196.423 & 3 & $3p^63d^2$ $^3F_3$ -- $3p^53d^3({\rm a}^2D)$ $^3F_4^{\rm o}$ &  5.6  &     4.7 \\ 
196.520 & 4 & 10.1 & 1.2 & 62.5 & 4.0 &     & \ion[Fe xiii] & 196.540 & 1 & $3s^23p^2$ $^1D_2$ -- $3s^23p3d$ $^1F_3$                    &  6.1  &    12.0 \\ 
196.664 & 1 & 79.4 & 1.9 & 87.9 & 2.0 & H & \ion[Fe xii] & 196.640 & 1 & $3s^23p^3$ $^2D_{5/2}^{\rm o}$ -- $3s^23p^2(^1D)3d$ $^2D_{5/2}$ &    6.1   &  30.7  \\ 
 &  &  &  &  &  &  & \ion[Fe viii] & 196.650 & 14 & $3p^63d$ $^2D_{3/2}$ -- $3p^53d^2(^1S)$ $^2P_{3/2}^{\rm o}$ &    5.7   &  30.5  \\ 
196.820 & 3 & 35.0 & 3.7 & 69.1 & 6.4 & E--G  \\ 
196.964 & 6 & 20.5 & 3.5 & 105.8 & 18.2 \\ 
197.193 & 7 & 8.3 & 2.4 & 79.8 & 23.4 \\ 
197.387 & 0 & 193.2 & 2.6 & 71.9 & 0.8 & D,H & \ion[Fe viii] & 197.362 & 14 & $3p^63d$ $^2D_{5/2}$ -- $3p^53d^2(^1S)$ $^2P_{3/2}^{\rm o}$ &    5.7   &  201.0  \\ 
197.871 & 1 & 188.1 & 12.4 & 76.2 & 2.4 & E & \ion[Fe ix] & 197.862 & 16 & $3s^23p^53d$ $^1P_{1}^{\rm o}$ -- $3s^2 3p^5 4p$ $^1S_{0}$ &    5.7   &  228.0  \\ 
198.091 & 6 & 7.5 & 1.7 & 62.4 & 14.4 \\ 
198.256 & 4 & 12.7 & 2.1 & 57.4 & 8.8 \\ 
198.411 & 8 & 13.8 & 2.7 & 96.9 & 19.0 \\ 
198.566 & 2 & 89.6 & 6.8 & 80.0 & 3.9 & G & \ion[S viii] & 198.550 & 15 & $2s^22p^5$ $^2P_{3/2}^{\rm o}$ -- $2s2p^6$ $^2S_{1/2}$ &    5.8   &  43.5  \\ 
 &  &  &  &  &  &  & \ion[Fe xi] & 198.546 & 1 & $3s^23p^4$ $^1D_{2}$ -- $3s^23p^3(^2D)3d$ $^3P_{1}$ &    6.0   &  20.0  \\ 
198.934 & 16 & 4.4 & 2.2 & 85.7 & 42.8 \\ 
199.200 & 12 & 6.3 & 1.5 & 85.0 & 2.7 \\ 
199.325 & 6 & 14.5 & 2.0 & 85.0 & 2.7 & E & \ion[Mn ix] & 199.319 & 1 & $3s^23p^5$ $^2P_{3/2}^{\rm o}$ -- $3s^23p^4(^1D)3d$ $^2S_{1/2}$ &    6.0   &  6.8  \\ 
199.613 & 3 & 34.4 & 2.8 & 85.0 & 2.7 E--F  \\ 
199.806 & 9 & 8.0 & 1.6 & 85.0 & 2.7 \\ 
200.003 & 3 & 47.7 & 3.4 & 85.0 & 2.7 & H & \ion[Fe xiii] & 200.022 & 1 & $3s^23p^2$ $^3P_{1}$ -- $3s^23p3d$ $^3D_{2}$ &    6.1   &  34.4  \\ 
 &  &  &  &  &  &  & \ion[Fe ix] & 199.986 & 2 & $3s^23p^53d$ $^1D_{2}^{\rm o}$ -- $3s^23p^4(^3P)3d^2$ $^3G_{3}$ &    5.7   &  11.8  \\
200.161 & 3 & 31.1 & 2.8 & 85.0 & 2.7 \\ 
200.384 & 5 & 16.8 & 1.9 & 85.0 & 2.7 \\ 
200.686 & 7 & 17.9 & 2.6 & 85.0 & 2.7 \\ 
200.785 & 3 & 45.7 & 3.8 & 85.0 & 2.7 \\ 
201.025 & 6 & 32.2 & 4.3 & 85.0 & 2.7 \\ 
201.113 & 3 & 83.1 & 6.0 & 85.0 & 2.7 & H & \ion[Fe xiii] & 201.128 & 1 & $3s^23p^2$ $^3P_{1}$ -- $3s^23p3d$ $^3D_{1}$ &    6.1   &  37.9  \\ 
201.515 & 8 & 16.3 & 2.7 & 86.9 & 1.6 \\ 
201.610 & 4 & 54.8 & 4.2 & 86.9 & 1.6 & G & \ion[Fe xi] & 201.577 & 1 & $3s^23p^4$ $^3P_{2}$ -- $3s^23p^3(^2P)3d$ $^3P_{2}$ &    6.0   &  26.7  \\ 
201.732 & 3 & 44.3 & 3.4 & 86.9 & 1.6 \\ 
201.887 & 2 & 47.7 & 2.9 & 72.9 & 4.3 & B,H & \ion[Fe vii] & 201.855 & 3 & $3p^63d^2$ $^3F_3$ -- $3p^53d^3(^2F)$ $^1G_4$ &  5.6  &     5.7 \\ 
202.038 & 1 & 190.9 & 5.2 & 88.7 & 2.1 & H & \ion[Fe xiii] & 202.044 & 1 & $3s^23p^2$ $^3P_{0}$ -- $3s^23p3d$ $^3P_{1}$ &    6.1   &  134.0  \\ 
202.344 & 13 & 9.2 & 3.9 & 55.6 & 23.5 \\ 
202.420 & 5 & 60.6 & 6.6 & 81.7 & 8.9 & G & \ion[Fe vii] & 202.378 & 3 & $3p^63d^2$ $^3F_4$ -- $3p^53d^3(^2F)$ $^1G_4$ &  5.6  & 1.2   \\ 
202.620 & 4 & 54.7 & 5.1 & 86.9 & 1.6 & E & \ion[S viii] & 202.605 & 15 & $2s^22p^5$ $^2P_{1/2}^{\rm o}$ -- $2s2p^6$ $^2S_{1/2}$ &    5.8   &  20.1  \\ 
 &  &  &  &  &  &  & \ion[Fe xi] & 202.628 & 1 & $3s^23p^4$ $^1D_{2}$ -- $3s^23p^3(^2D)3d$ $^3P_{2}$ &    6.0   &  5.8  \\ 
202.708 & 6 & 38.0 & 4.2 & 86.9 & 1.6 & G & \ion[Fe xi] & 202.706 & 1 & $3s^23p^4$ $^1D_{2}$ -- $3s^23p^3(^2D)3d$ $^1P_{1}$ &    6.0   &  48.0  \\ 
202.857 & 1 & 194.5 & 10.9 & 86.9 & 1.6 & C & \ion[Cr vii] & 202.828 & 11 & $3s^2 3p^6$ $^1S_{0}$ -- $3s^2 3p^5 3d$ $^1P_{1}^{\rm o}$ &    5.7   &  249.0  \\ 
203.064 & 13 & 12.2 & 3.5 & 86.9 & 1.6 \\ 
203.138 & 11 & 18.4 & 3.5 & 86.9 & 1.6 & I & \ion[Fe xiii] & 203.164 & 1 & $3s^23p^2$ $^3P_{1}$ -- $3s^23p3d$ $^3P_{0}$ &    6.1   &  17.6  \\ 
203.249 & 8 & 11.7 & 1.9 & 86.9 & 1.6 \\ 
203.634 & 12 & 15.7 & 3.9 & 100.2 & 5.1 \\ 
203.737 & 9 & 72.4 & 12.3 & 100.2 & 5.1 & H & \ion[Fe xii] & 203.728 & 1 & $3s^23p^3$ $^2D_{5/2}^{\rm o}$ -- $3s^23p^2(^1S)3d$ $^2D_{5/2}$ &    6.1   &  28.2  \\ 
203.824 & 4 & 141.1 & 12.7 & 100.2 & 5.1 & H & \ion[Fe xiii] & 203.828 & 1 & $3s^23p^2$ $^3P_{2}$ -- $3s^23p3d$ $^3D_{3}$ &    6.1   &  115.0  \\ 
 &  &  &  &  &  &  & \ion[Fe xiii] & 203.797 & 1 & $3s^23p^2$ $^3P_{2}$ -- $3s^23p3d$ $^3D_{2}$ &    6.1   &  48.0  \\ 
203.998 & 7 & 15.4 & 1.9 & 100.2 & 5.1 \\ 
204.173 & 8 & 25.2 & 3.4 & 100.2 & 5.1 \\ 
204.274 & 26 & 19.4 & 7.9 & 100.2 & 5.1 & H--I & \ion[Fe xiii] & 204.263 & 1 & $3s^23p^2$ $^3P_{1}$ -- $3s^23p3d$ $^1D_{2}$ &    6.1   &  14.0  \\ 
204.340 & 32 & 11.6 & 9.1 & 100.2 & 5.1 \\ 
204.468 & 10 & 10.6 & 1.8 & 100.2 & 5.1 \\ 
204.722 & 1 & 211.9 & 11.9 & 72.7 & 1.5 \\ 
204.911 & 2 & 38.0 & 2.9 & 72.7 & 1.5 & H--I & \ion[Fe xiii] & 204.945 & 1 & $3s^23p^2$ $^3P_{2}$ -- $3s^23p3d$ $^3D_{1}$ &    6.1   &  11.7  \\ 
205.053 & 1 & 131.2 & 7.5 & 72.7 & 1.5 & C & \ion[Cr viii] & 205.010 & 13 & $3s^23p^5$ $^2P_{3/2}^{\rm o}$ -- $3s^23p^4(^3P)3d$ $^2D_{5/2}$ &    5.7   &  48.4  \\ 
205.605 & 7 & 7.8 & 1.5 & 72.7 & 1.5 \\ 
205.717 & 2 & 39.1 & 3.1 & 72.7 & 1.5 &  & \ion[Cr viii] & 205.650 & 13 & $3s^23p^5$ $^2P_{1/2}^{\rm o}$ -- $3s^23p^4(^3P)3d$ $^2D_{3/2}$ &    5.7   &  10.5  \\ 
206.161 & 4 & 15.5 & 1.8 & 72.7 & 1.5 \\ 
206.269 & 4 & 19.7 & 2.2 & 72.7 & 1.5 \\ 
206.362 & 4 & 17.1 & 2.0 & 72.7 & 1.5 &  & \ion[Fe xii] & 206.368 & 1 & $3s^23p^3$ $^2D_{3/2}^{\rm o}$ -- $3s^23p^2(^1S)3d$ $^2D_{3/2}$ &    6.1   &  6.5  \\ 
206.775 & 2 & 50.5 & 4.0 & 71.4 & 2.7 &  & \ion[Fe viii] & 206.753 & 2 & $3p^63d$ $^2D_{5/2}$ -- $3p^53d^2(^1G)$ $^2G_{7/2}$ &    5.6   &  65.3  \\ 
207.031 & 5 & 19.7 & 2.3 & 71.4 & 2.7 \\ 
207.141 & 2 & 198.5 & 14.1 & 71.4 & 2.7 \\ 
207.215 & 4 & 51.5 & 7.8 & 71.4 & 2.7 \\ 
207.458 & 2 & 98.5 & 6.7 & 82.3 & 3.1 \\ 
207.743 & 3 & 89.3 & 8.1 & 72.3 & 6.0 & B--C & \ion[Fe vii] & 207.712 & 3 & $3p^63d^2$ $^3F_2$ -- $3p^53d^3(^2F)$ $^3G_3^{\rm o}$ &  5.6  &    21.4 \\ 
207.948 & 3 & 22.8 & 2.8 & 69.0 & 6.5 &  & \ion[Ni xi] & 207.922 & 1 & $3s^23p^6$ $^1S_{0}$ -- $3s^23p^53d$ $^3P_{2}$ &    6.0   &  17.3  \\ 
208.199 & 11 & 9.7 & 5.6 & 48.8 & 28.2 &  & \ion[Fe vii] & 208.167 & 3 & $3p^63d^2$ $^3F_3$ -- $3p^53d^3(^2F)$ $^3G_3^{\rm o}$ &  5.6  &  3.0  \\ 
208.679 & 2 & 71.4 & 5.6 & 79.2 & 3.9 &  & \ion[Cr viii] & 208.630 & 13 & $3s^23p^5$ $^2P_{3/2}^{\rm o}$ -- $3s^23p^4(^3P)3d$ $^2P_{3/2}$ &    5.7   &  23.2  \\ 
208.846 & 2 & 99.0 & 7.3 & 71.8 & 2.9 \\ 
209.453 & 2 & 120.5 & 8.3 & 85.9 & 3.5 \\ 
209.659 & 6 & 93.9 & 10.1 & 132.2 & 14.2 & I & \ion[Fe xiii] & 209.621 & 1 & $3s^23p^2$ $^3P_{1}$ -- $3s^23p3d$ $^3P_{2}$ &    6.1   &  23.8  \\ 
209.760 & 3 & 40.0 & 5.3 & 46.6 & 6.2 \\ 
209.936 & 2 & 113.9 & 7.8 & 85.9 & 3.5 & H--I & \ion[Fe xiii] & 209.919 & 1 & $3s^23p^2$ $^3P_{2}$ -- $3s^23p3d$ $^3P_{1}$ &    6.1   &  20.1  \\ 
210.665 & 2 & 52.4 & 4.3 & 57.7 & 3.3 \\ 
211.306 & 2 & 124.0 & 8.4 & 131.4 & 5.7 & H--I & \ion[Fe xiv] & 211.318 & 1 & $3s^23p$ $^2P_{1/2}^{\rm o}$ -- $3s^23d$ $^2D_{3/2}$ &    6.2   &  136.0  \\ 
211.484 & 4 & 16.3 & 2.4 & 55.0 & 8.2 &  & \ion[Ni xi] & 211.428 & 1 & $3s^23p^6$ $^1S_{0}$ -- $3s^23p^53d$ $^3P_{1}$ &    6.0   &  6.7  \\ 
 &  &  &  &  &  &  & \ion[Cr viii] & 211.420 & 13 & $3s^23p^5$ $^2P_{1/2}^{\rm o}$ -- $3s^23p^4(^3P)3d$ $^2P_{1/2}$ &    5.7   &  4.4  \\ 
 & \\
245.970 & 10 & 35.5 & 16.6 & 80.3 & 2.7 \\ 
246.036 & 2 & 457.7 & 29.6 & 80.3 & 2.7 & C & \ion[Si vi] & 246.004 & 8 & $2s^22p^5$ $^2P_{3/2}$ -- $2s2p^6$ $^2S_{1/2}$ &  5.6 &  543.0 \\ 
246.181 & 3 & 27.3 & 2.2 & 80.3 & 2.7 & H--I & \ion[Fe xiii] & 246.211 & 1 & $3s^23p^2$ $^3P_{1}$ -- $3s3p^3$ $^3S_{1}^{\rm o}$ &    6.1   &  36.8  \\ 
 &  &  &  &  &  &  & \ion[Si vii] & 246.124 & 6 & $2s^22p^4$ $^1S_{0}$ -- $2s2p^5$ $^1P_{1}^{\rm o}$ &  5.7   &  5.0  \\ 
247.018 & 2 & 20.1 & 2.1 & 49.5 & 3.7 \\ 
247.426 & 2 & 32.7 & 2.9 & 62.4 & 3.8 & E & \ion[Al viii] & 247.404 & 7 & $2p^2$ $^3P_0$ -- $2s2p^3$ $^3S_1^{\rm o}$ &  6.0  &  4.6 \\ 
248.489 & 1 & 105.9 & 6.8 & 70.6 & 2.2 & A & \ion[O v] & 248.460 & 1 & $2s2p$ $^1P_{1}^{\rm o}$ -- $2s3s$ $^1S_{0}$ &    5.6   &  126.0  \\ 
 &  &  &  &  &  &  & \ion[Al viii] & 248.458 & 7 & $2p^2$ $^3P_1$ -- $2s2p^3$ $^3S_1^{\rm o}$ &  6.0  &  13.8 \\ 
248.668 & 3 & 34.7 & 3.9 & 61.7 & 6.9 & A--B & \ion[Fe vii] & 248.641 & 2 & $3p^63d^2$ $^3F_4$ -- $3p^53d^3(^4F)$ $^5F_4^{\rm o}$ &  5.6  &    8.4 \\ 
249.163 & 1 & 212.8 & 13.5 & 73.2 & 1.9 & C,M & \ion[Si vi] & 249.124 & 8 & $2s^22p^5$ $^2P_{1/2}$ -- $2s2p^6$ $^2S_{1/2}$ &  5.6 &  257.0 \\ 
249.329 & 3 & 70.4 & 5.4 & 87.2 & 6.1 & H--I & \ion[Fe vii] & 249.302 & 2 & $3p^63d^2$ $^3F_4$ -- $3p^53d^3(^4F)$ $^5F_5^{\rm o}$ &  5.6  &    14.2 \\ 
250.155 & 2 & 24.7 & 2.4 & 60.1 & 4.1 & E & \ion[Al viii] & 250.139 & 7 & $2p^2$ $^3P_2$ -- $2s2p^3$ $^3S_1^{\rm o}$ &  6.0  &  23.2  \\ 
250.514 & 3 & 5.3 & 1.0 & 35.7 & 5.9 &  & \ion[Si viii] & 250.465 & 5 & $2s^22p^3$ $^2P_{1/2}^{\rm o}$ -- $2s2p^4$ $^2S_{1/2}$ &  6.0 &  3.4 \\ 
250.823 & 5 & 6.9 & 1.1 & 61.1 & 9.4 &  & \ion[Si viii] & 250.807 & 5 & $2s^22p^3$ $^2P_{3/2}^{\rm o}$ -- $2s2p^4$ $^2S_{1/2}$ &  6.0 &  5.2 \\ 
251.932 & 3 & 44.1 & 3.0 & 113.8 & 5.3 & H--I & \ion[Fe xiii] & 251.956 & 1 & $3s^23p^2$ $^3P_{2}$ -- $3s3p^3$ $^3S_{1}^{\rm o}$ &    6.1   &  70.9  \\ 
252.918 & 4 & 12.9 & 1.7 & 88.8 & 10.0 \\ 
253.555 & 3 & 19.0 & 2.7 & 55.2 & 7.8 & A--D & \ion[Fe vii] & 253.528 & 2 & $3p^63d^2$ $^3F_3$ -- $3p^53d^3(^4F)$ $^5D_3^{\rm o}$ &  5.6  &     5.2 \\ 
253.771 & 4 & 19.9 & 1.9 & 109.1 & 9.1 & H & \ion[Si x] & 253.788 & 1 & $2s^22p$ $^2P_{1/2}^{\rm o}$ -- $2s2p^2$ $^2P_{3/2}$ &    6.1   &  26.1  \\ 
253.981 & 1 & 285.9 & 7.1 & 70.1 & 1.6 & C--D & \ion[Fe viii] & 253.953 & 1 & $3p^63d$ $^2D_{5/2}$ -- $3p^53d^2(^3F)$ $^4D_{7/2}$ &    5.6   &  223.0  \\ 
254.085 & 3 & 49.4 & 4.6 & 68.7 & 6.4 & B & \ion[Fe vii] & 254.057 & 2 & $3p^63d^2$ $^3F_4$ -- $3p^53d^3(^4F)$ $^5D_4^{\rm o}$ &  5.6  &    11.4 \\ 
254.223 & 2 & 22.0 & 2.1 & 60.9 & 4.2 \\ 
254.590 & 4 & 11.9 & 1.5 & 62.1 & 7.5 \\ 
254.706 & 2 & 36.7 & 3.1 & 71.6 & 4.2 \\ 
255.132 & 1 & 112.9 & 4.5 & 66.1 & 2.1 & C--D & \ion[Fe viii] & 255.103 & 1 & $3p^63d$ $^2D_{3/2}$ -- $3p^53d^2(^3F)$ $^4D_{3/2}$ &    5.6   &  129.0  \\ 
255.373 & 1 & 193.9 & 5.5 & 72.2 & 1.6 & C--D & \ion[Fe viii] & 255.344 & 1 & $3p^63d$ $^2D_{5/2}$ -- $3p^53d^2(^3F)$ $^4D_{5/2}$ &    5.6   &  127.0  \\ 
255.707 & 2 & 56.4 & 3.3 & 66.1 & 3.4 & C & \ion[Fe viii] & 255.678 & 1 & $3p^63d$ $^2D_{3/2}$ -- $3p^53d^2(^3F)$ $^4D_{1/2}$ &    5.6   &  89.5  \\ 
256.045 & 5 & 8.6 & 1.4 & 78.1 & 13.0 \\ 
256.345 & 1 & 608.0 & 27.9 & 95.9 & 1.5 & A & \ion[He ii] & 256.317 & 1 & $1s$ $^2S_{1/2}$ -- $3p$ $^2P_{3/2}^{\rm o}$ &    4.9   &  1270.0  \\ 
 &  &  &  &  &  &  & \ion[He ii] & 256.318 & 1 & $1s$ $^2S_{1/2}$ -- $3p$ $^2P_{1/2}^{\rm o}$ &    4.9   &  632.0  \\ 
 &  &  &  &  &  &  & \ion[Si x] & 256.366 & 1 & $2s^22p$ $^2P_{1/2}^{\rm o}$ -- $2s2p^2$ $^2P_{1/2}$ &    6.1   &  56.6  \\ 
256.466 & 4 & 65.4 & 6.7 & 95.9 & 1.5 & G & \ion[Fe x] & 256.398 & 1 & $3s^23p^5$ $^2P_{3/2}^{\rm o}$ -- $3s^23p^4(^3P)3d$ $^4D_{3/2}$ &    6.0   &  22.9  \\ 
 &  &  &  &  &  &  & \ion[Fe xii] & 256.410 & 1 & $3s^23p^3$ $^2D_{5/2}^{\rm o}$ -- $3s^23p^2(^3P)3d$ $^4F_{7/2}$ &    6.1   &  14.5  \\ 
 &  &  &  &  &  &  & \ion[Fe xiii] & 256.422 & 1 & $3s^23p^2$ $^1D_{2}$ -- $3s3p^3$ $^1P_{1}^{\rm o}$ &    6.1   &  13.6  \\ 
256.922 & 3 & 35.1 & 2.6 & 95.9 & 1.5 \\ 
257.182 & 10 & 17.0 & 3.8 & 83.9 & 1.4 &  & \ion[S x] & 257.147 & 1 & $2s^22p^3$ $^4S_{3/2}$ -- $2s2p^4$ $^4P_{1/2}$ &  6.1   &   7.9 \\ 
257.275 & 1 & 396.4 & 7.3 & 83.9 & 1.4 & F & \ion[Fe x] & 257.259 & 1 & $3s^23p^5$ $^2P_{3/2}^{\rm o}$ -- $3s^23p^4(^3P)3d$ $^4D_{5/2}$ &    6.0   &  55.6  \\ 
 &  &  &  &  &  &  & \ion[Fe x] & 257.263 & 1 & $3s^23p^5$ $^2P_{3/2}^{\rm o}$ -- $3s^23p^4(^3P)3d$ $^4D_{7/2}$ &    6.0   &  126.0  \\ 
257.420 & 6 & 17.7 & 2.0 & 83.9 & 1.4 \\ 
257.545 & 3 & 41.8 & 2.4 & 83.9 & 1.4 & G--H & \ion[Fe xi] & 257.547 & 1 & $3s^23p^4$ $^3P_{2}$ -- $3s^23p^3(^4S)3d$ $^5D_3^{\rm o}$ &  6.0  &  12.3  \\ 
257.768 & 5 & 17.4 & 2.0 & 83.9 & 1.4 & G--H & \ion[Fe xi] & 257.772 & 1 & $3s^23p^4$ $^3P_{2}$ -- $3s^23p^3(^4S)3d$ $^5D_2^{\rm o}$ &  6.0  &  6.3  \\ 
258.089 & 3 & 21.2 & 2.1 & 85.4 & 7.2 &  & \ion[Si ix] & 258.082 & 7 & $2s^22p^2$ $^1D_2$ -- $2s2p^3$ $^1D_2^{\rm o}$ &  6.0   &   20.3 \\ 
258.369 & 2 & 109.4 & 6.7 & 95.5 & 3.0 & G--H & \ion[Si x] & 258.371 & 1 & $2s^22p$ $^2P_{3/2}^{\rm o}$ -- $2s2p^2$ $^2P_{3/2}$ &    6.1   &  136.0  \\ 
258.636 & 4 & 5.1 & 1.0 & 51.6 & 10.0 &      & \ion[Cr vii] & 258.655 & 11 & $3s^2 3p^6$ $^1S_{0}$ -- $3s^2 3p^5 3d$ $^3D_{2}^{\rm o}$ &    5.7   &  1.5  \\ 
259.226 & 2 & 13.0 & 1.6 & 55.4 & 4.9 & A--D & \ion[Cr vii] & 259.181 & 11 & $3s^2 3p^6$ $^1S_{0}$ -- $3s^2 3p^5 3d$ $^3D_{1}^{\rm o}$ &    5.7   &  3.5  \\ 
 &  &  &  &  &  &  & \ion[Al vii] & 259.196 & 5 & $2s^2 2p^3$ $^2P_{3/2}^{\rm o}$ -- $2s 2p^4$ $^2P_{1/2}$ &    5.7   &  2.3  \\ 
259.515 & 4 & 27.7 & 2.2 & 127.9 & 8.2 & H & \ion[S x] & 259.497 & 1 & $2s^22p^3$ $^4S_{3/2}^{\rm o}$ -- $2s2p^4$ $^4P_{3/2}$ &    6.1   &  15.3  \\ 
259.985 & 2 & 27.9 & 2.4 & 72.6 & 4.2 \\ 
260.141 & 3 & 4.8 & 1.0 & 40.2 & 6.9 \\ 
260.295 & 4 & 16.7 & 1.8 & 111.2 & 10.1 &  & \ion[O iv] & 260.389 & 1 & $2s 2p^2$ $^2D_{5/2}$ -- $2s 2p (^3P)  3d$ $^2F_{7/2}^{\rm o}$ &    5.2   &  16.9  \\ 
260.707 & 3 & 19.8 & 1.9 & 67.7 & 6.1 & B--C & \ion[Fe vii] & 260.678 & 2 & $3p^63d^2$ $^3P_2$ -- $3p^53d^3(^4F)$ $^5F_3^{\rm o}$ &  5.6  &     5.3 \\ 
261.053 & 2 & 47.9 & 3.4 & 91.0 & 4.0 & H & \ion[Si x] & 261.044 & 1 & $2s^22p$ $^2P_{3/2}^{\rm o}$ -- $2s2p^2$ $^2P_{1/2}$ &    6.1   &  50.6  \\ 
261.242 & 4 & 4.6 & 0.9 & 50.6 & 10.0 &  & \ion[Al vii] & 261.209 & 1 & $2s^22p^3$ $^2P_{3/2}^{\rm o}$ -- $2s2p^4$ $^2P_{3/2}$ &    5.7   &  4.8  \\ 
261.354 & 6 & 5.0 & 1.1 & 70.1 & 14.9 &      & \ion[Cr vii] & 261.314 & 11 & $3s^2 3p^6$ $^1S_{0}$ -- $3s^2 3p^5 3d$ $^1D_{2}^{\rm o}$ &    5.7   &  0.6 \\ 
261.728 & 6 & 7.5 & 1.3 & 94.8 & 15.0 \\ 
262.303 & 6 & 4.0 & 0.9 & 58.9 & 13.3 \\ 
262.696 & 17 & 2.1 & 1.0 & 87.0 & 39.3 \\ 
262.988 & 4 & 8.4 & 1.2 & 70.8 & 9.3 & L--M & \ion[Fe xvi] & 262.976 & 1 & $3p$ $^2P_{3/2}^{\rm o}$ -- $3d$ $^2D_{5/2}$ &    6.4   &  5.9  \\ 
263.200 & 4 & 10.8 & 1.4 & 76.0 & 8.6 &  & \ion[Mn viii] & 263.163 & 12 & $3s^2 3p^6$ $^1S_{0}$ -- $3s^2 3p^5 3d$ $^3P_{2}^{\rm o}$ &    5.7 & 4.0 \\ 
264.239 & 3 & 21.9 & 2.1 & 97.6 & 7.4 & H & \ion[S x] & 264.231 & 1 & $2s^22p^3$ $^4S_{3/2}^{\rm o}$ -- $2s2p^4$ $^4P_{5/2}$ &    6.1   &  22.3  \\ 
264.375 & 7 & 3.2 & 1.0 & 54.9 & 16.7 \\ 
264.630 & 20 & 5.2 & 2.0 & 101.9 & 39.2 \\ 
264.780 & 3 & 68.4 & 4.4 & 115.1 & 6.1 & I & \ion[Fe xiv] & 264.790 & 1 & $3s^23p$ $^2P_{3/2}^{\rm o}$ -- $3s3p^2$ $^2P_{3/2}$ &    6.2   &  68.2  \\ 
265.738 & 3 & 11.6 & 1.3 & 64.8 & 7.2 & B,L & \ion[Fe vii] & 265.697 & 3 & $3p^63d^2$ $^1S_0$ -- $3p^63d4p$ $^1P_1^{\rm o}$ &  5.6  &     4.9 \\ 
266.106 & 5 & 14.7 & 1.6 & 87.4 & 3.6 \\ 
266.208 & 5 & 13.1 & 1.5 & 87.4 & 3.6 \\ 
266.531 & 8 & 9.1 & 1.5 & 87.4 & 3.6 \\ 
266.623 & 5 & 15.1 & 1.8 & 87.4 & 3.6 \\ 
267.245 & 6 & 13.7 & 2.8 & 70.4 & 6.8 & B & \ion[Fe vii] & 267.231 & 2 & $3p^63d^2$ $^3P_2$ -- $3p^53d^3(^4F)$ $^5D_3^{\rm o}$ &  5.6  & 1.4  \\
 &  &  &  &  &  &  & \ion[Fe vii] & 267.250 & 2 & $3p^63d^2$ $^3P_2$ -- $3p^53d^3(^4F)$ $^5D_1^{\rm o}$ &  5.6  & 1.7  \\
267.303 & 4 & 19.8 & 2.4 & 70.4 & 6.8 & B & \ion[Fe vii] & 267.274 & 2 & $3p^63d^2$ $^3P_2$ -- $3p^53d^3(^4F)$ $^5D_2^{\rm o}$ &  5.6  & 4.3  \\
268.043 & 7 & 6.0 & 1.0 & 87.4 & 3.6 \\ 
268.228 & 8 & 5.1 & 1.0 & 87.4 & 3.6 \\ 
269.020 & 1 & 249.7 & 15.0 & 83.6 & 2.0 & C & \ion[Mg vi] & 268.991 & 5 & $2s^22p^3$ $^2D_{3/2}^{\rm o}$ -- $2s2p^4$ $^2P_{1/2}$ &    5.7   &  213.0  \\ 
269.561 & 6 & 8.4 & 1.2 & 83.6 & 2.0 \\ 
269.817 & 14 & 2.8 & 1.0 & 83.6 & 2.0 \\ 
270.426 & 1 & 522.2 & 30.7 & 83.9 & 2.2 & C & \ion[Mg vi] & 270.390 & 5 & $2s^22p^3$ $^2D_{5/2}^{\rm o}$ -- $2s2p^4$ $^2P_{3/2}$ &    5.7   &  370.0  \\ 
 &  &  &  &  &  &  & \ion[Mg vi] & 270.400 & 5 & $2s^22p^3$ $^2D_{3/2}^{\rm o}$ -- $2s2p^4$ $^2P_{3/2}$ &    5.7   &  49.5  \\ 
270.539 & 8 & 17.4 & 3.5 & 67.2 & 13.5 & I & \ion[Fe xiv] & 270.522 & 1 & $3s^23p$ $^2P_{3/2}^{\rm o}$ -- $3s3p^2$ $^2P_{1/2}$ &    6.2   &  38.9  \\ 
271.068 & 6 & 17.6 & 1.9 & 124.1 & 13.1 & A & \ion[O v] & 270.978 & 1 & $2p^2$ $^3P_{2}$ -- $2s3p$ $^3P_{2}^{\rm o}$ &    5.6   &  9.3  \\ 
 &  &  &  &  &  &  & \ion[O v] & 271.035 & 1 & $2p^2$ $^3P_{2}$ -- $2s3p$ $^3P_{1}^{\rm o}$ &    5.6   &  1.6  \\ 
271.729 & 2 & 24.7 & 1.7 & 79.8 & 5.1 & B & \ion[Fe vii] & 271.699 & 2 & $3p^63d^2$ $^3P_2$ -- $3p^53d^3(^4P)$ $^5S_2^{\rm o}$ &  5.6  &     5.3 \\ 
271.990 & 3 & 51.2 & 3.8 & 113.4 & 7.3 & H & \ion[Si x] & 272.006 & 1 & $2s^22p$ $^2P_{1/2}^{\rm o}$ -- $2s2p^2$ $^2S_{1/2}$ &    6.1   &  38.1  \\ 
272.153 & 7 & 16.2 & 2.5 & 99.7 & 15.6 \\ 
272.339 & 6 & 11.2 & 2.0 & 94.6 & 16.8 \\ 
272.685 & 1 & 235.7 & 12.6 & 84.6 & 1.6 & D & \ion[Si vii] & 272.647 & 6 & $2s^22p^4$ $^3P_{2}$ -- $2s2p^5$ $^3P_{1}^{\rm o}$ &  5.7 &  214.0 \\ 
274.119 & 7 & 15.4 & 2.8 & 84.6 & 1.6 \\ 
274.218 & 2 & 217.2 & 11.4 & 84.6 & 1.6 & C,I & \ion[Si vii] & 274.180 & 6 & $2s^22p^4$ $^3P_{1}$ -- $2s2p^5$ $^3P_{0}^{\rm o}$ &  5.7 &  142.0 \\ 
 &  &  &  &  &  &  & \ion[Fe xiv] & 274.204 & 1 & $3s^23p$ $^2P_{1/2}^{\rm o}$ -- $3s3p^2$ $^2S_{1/2}$ &  6.2 &  74.9 \\ 
275.394 & 1 & 699.4 & 37.9 & 90.5 & 1.5 & C & \ion[Si vii] & 275.361 & 6 & $2s^22p^4$ $^3P_{2}$ -- $2s2p^5$ $^3P_{2}^{\rm o}$ &  5.7 &  706.0 \\ 
275.715 & 1 & 133.3 & 3.1 & 78.0 & 1.4 & C & \ion[Si vii] & 275.675 & 6 & $2s^22p^4$ $^3P_1$ -- $2s2p^5$ $^3P_1^{\rm o}$ &  5.7 &  123.0 \\ 
276.174 & 1 & 142.9 & 2.8 & 79.2 & 1.0 & C & \ion[Mg vii] & 276.138 & 7 & $2s^22p^2$ $^3P_0$ -- $2s2p^3$ $^3S_1^{\rm o}$ &  5.7 &  100.0 \\ 
276.625 & 1 & 225.0 & 4.4 & 76.5 & 1.1 & B & \ion[Mg v] & 276.579 & 6 & $2s^22p^4$ $^1D_2$ -- $2s2p^5$ $^1P_1^{\rm o}$ &  5.6 &  245.0 \\ 
276.873 & 2 & 140.9 & 6.4 & 100.2 & 2.4 &  & \ion[Si viii] & 276.850 & 5 & $2s^22p^3$ $^2D_{3/2}^{\rm o}$ -- $2s2p^4$ $^2D_{3/2}$ &  6.0 &  85.2 \\ 
276.881 & 2 & 8.0 & 0.3 & 100.2 & 2.4 &  & \ion[Si viii] & 276.865 & 5 & $2s^22p^3$ $^2D_{3/2}^{\rm o}$ -- $2s2p^4$ $^2D_{5/2}$ &  6.0 &  4.7 \\ 
276.891 & 1 & 174.7 & 4.1 & 78.0 & 1.4 &  & \ion[Si vii] & 276.850 & 6 & $2s^22p^4$ $^3P_0$ -- $2s2p^5$ $^3P_1^{\rm o}$ &  5.7 &  160.0 \\ 
277.023 & 1 & 427.4 & 8.2 & 79.2 & 1.0 &  & \ion[Mg vii] & 276.993 & 7 & $2s^22p^2$ $^3P_1$ -- $2s2p^3$ $^3S_1^{\rm o}$ &  5.7 &  299.0 \\ 
277.066 & 2 & 12.3 & 0.6 & 100.2 & 2.4 &  & \ion[Si viii] & 277.042 & 5 & $2s^22p^3$ $^2D_{5/2}^{\rm o}$ -- $2s2p^4$ $^2D_{3/2}$ &  6.0 &  7.4 \\ 
277.074 & 2 & 201.0 & 8.7 & 100.2 & 2.4 &  & \ion[Si viii] & 277.057 & 5 & $2s^22p^3$ $^2D_{5/2}^{\rm o}$ -- $2s2p^4$ $^2D_{5/2}$ &  6.0 &  119.0 \\ 
277.255 & 3 & 31.7 & 2.3 & 97.9 & 7.0 & H & \ion[Si x] & 277.278 & 1 & $2s^22p$ $^2P_{3/2}^{\rm o}$ -- $2s2p^2$ $^2S_{1/2}$ &  6.1 &  31.2 \\ 
277.625 & 5 & 11.6 & 1.6 & 86.5 & 11.3 \\ 
278.426 & 1 & 750.2 & 17.5 & 87.3 & 1.7 & D & \ion[Mg vii] & 278.393 & 7 & $2s^22p^2$ $^3P_2$ -- $2s2p^3$ $^3S_1^{\rm o}$ &  5.7 &  502.0 \\ 
278.484 & 2 & 236.1 & 26.4 & 87.3 & 1.7 & D & \ion[Si vii] & 278.449 & 6 & $2s^22p^4$ $^3P_1$ -- $2s2p^5$ $^3P_2^{\rm o}$ &  5.7 &  226.0 \\ 
278.731 & 3 & 29.4 & 2.8 & 84.4 & 5.8 & B & \ion[Al v] & 278.694 & 10 & $2s^2 2p^5$ $^2P_{3/2}^{\rm o}$ -- $2s 2p^6$ $^2S_{1/2}$ &    5.6   &  20.9  \\ 
 &  &  &  &  &  &  & \ion[Ni xi] & 278.684 & 17 & $3s^23p^53d$ $^3F_{4}$ -- $3s3p^63d$ $^3D_3$ & 5.9 & 3.0 \\
278.898 & 5 & 6.4 & 1.2 & 59.6 & 10.9 \\ 
279.670 & 3 & 14.7 & 1.8 & 72.7 & 7.5 & A & \ion[O iv] & 279.631 & 1 & $2s^22p$ $^2P_{1/2}^{\rm o}$ -- $2s^23s$ $^2S_{1/2}$ &    5.2   &  22.0  \\ 
279.973 & 2 & 43.1 & 3.4 & 73.0 & 3.7 & A & \ion[O iv] & 279.933 & 1 & $2s^22p$ $^2P_{3/2}^{\rm o}$ -- $2s^23s$ $^2S_{1/2}$ &    5.2   &  44.1  \\ 
280.759 & 1 & 220.0 & 12.9 & 83.4 & 1.9 & D & \ion[Mg vii] & 280.722 & 7 & $2s^22p^2$ $^1D_2$ -- $2s2p^3$ $^1P_1^{\rm o}$ &  5.7 &  176.0 \\ 
281.438 & 3 & 10.5 & 1.4 & 54.7 & 6.2 & A--D,I & \ion[Al v] & 281.394 & 10 & $2s^2 2p^5$ $^2P_{1/2}^{\rm o}$ -- $2s 2p^6$ $^2S_{1/2}$ &    5.6   &  10.0  \\ 
282.443 & 4 & 14.2 & 1.7 & 79.3 & 7.7 &  & \ion[Al ix] & 282.421 & 1 & $2s^22p$ $^2P_{1/2}^{\rm o}$ -- $2s2p^2$ $^2P_{1/2}$ &    6.0   &  10.0  \\ 
283.981 & 9 & 16.3 & 3.5 & 84.5 & 3.0 \\ 
284.063 & 6 & 50.3 & 4.8 & 84.5 & 3.0 & G--H & \ion[Al ix] & 284.025 & 1 & $2s^22p$ $^2P_{3/2}^{\rm o}$ -- $2s2p^2$ $^2P_{3/2}$ &    6.0   &  27.4  \\ 
284.173 & 2 & 161.7 & 10.2 & 84.5 & 3.0 & L & \ion[Fe xv] & 284.163 & 1 & $3s^2$ $^1S_{0}$ -- $3s3p$ $^1P_{1}^{\rm o}$ &    6.3   &  281.0  \\ 
286.375 & 5 & 8.4 & 1.3 & 69.2 & 9.2 &  & \ion[Al ix] & 286.376 & 1 & $2s^22p$ $^2P_{3/2}^{\rm o}$ -- $2s2p^2$ $^2P_{1/2}$ &    6.0   &  7.9  \\ 
289.727 & 4 & 38.0 & 5.3 & 74.1 & 10.4 & A--D & \ion[Fe vii] & 289.678 & 2 & $3p^63d^2$ $^3F_2$ -- $3p^63d4s$ $^3D_2$ &  5.6  &    12.8 \\ 
289.884 & 3 & 36.6 & 4.8 & 57.2 & 7.5 & A--D & \ion[Fe vii] & 289.831 & 2 & $3p^63d^2$ $^3F_3$ -- $3p^63d4s$ $^3D_3$ &  5.6  &    12.6 \\ 
290.345 & 3 & 69.6 & 6.2 & 72.2 & 6.2 & B & \ion[Fe vii] & 290.307 & 2 & $3p^63d^2$ $^3F_2$ -- $3p^63d4s$ $^3D_1$ &  5.6  &    17.6 \\ 
290.704 & 10 & 34.1 & 7.5 & 80.1 & 4.7 &  & \ion[Si ix] & 290.687 & 7 & $2s^22p^2$ $^3P_0$ -- $2s2p^3$ $^3P_1^{\rm o}$ &  6.0   &   34.8 \\ 
290.791 & 2 & 178.8 & 10.4 & 80.1 & 4.7 & B & \ion[Fe vii] & 290.717 & 2 & $3p^63d^2$ $^3F_3$ -- $3p^63d4s$ $^3D_2$ &  5.6  &    21.1 \\ 
 &  &  &  &  &  &  & \ion[Fe vii] & 290.756 & 2 & $3p^63d^2$ $^3F_4$ -- $3p^63d4s$ $^3D_3$ &  5.6  &    45.2 \\ 

\enddata
\end{deluxetable}

\begin{table}[h]
\begin{center}
\caption{Reference wavelength sources.\label{tbl.ref-wvl}}
\begin{tabular}{ll}
\tableline\tableline
Index & Reference \\
\tableline
1  &CHIANTI 6.0 \citet{dere09}   \\
2  &\citet{young09a} \\
3  &\citet{ekberg81}   \\
4  &\citet{edlen34}   \\
5  &\citet{edlen84}   \\
6  &\citet{edlen83}   \\
7  &\citet{edlen85}   \\
8  &\citet{artru77}   \\
9  &\citet{edlen79}   \\
10 &\citet{artru74}   \\
11 &\citet{ekberg76}   \\
12 &\citet{smitt83}   \\
13 &\citet{fawcett66}   \\
14 &\citet{ramonas80} \\
15 & \citet{robinson37} \\
16 & \citet{young09} \\
17 &Present work   \\
\tableline
\end{tabular}
\end{center}
\end{table}

\begin{table}[h]
\begin{center}
\caption{Emission line velocities.\label{tbl.vel}}
\begin{tabular}{llllll}
\tableline\tableline
Index & $T_{\rm eff}$ & $\lambda$ &$v_{\rm ref}$ & $v_{\rm NIST}$ & $\sigma_v$ \\
\tableline
\ionx{O}{iv} & 5.21&  279.670 &    41.8 &    41.8 &     3.2 \\
 & &  279.973 &    42.8 &    42.8 &     2.1 \\
\ionx{O}{v} & 5.49&  192.931 &    42.0 &    38.9 &     1.6 \\
 & &  248.489 &    35.0 &    36.2 &     1.2 \\
\ionx{Al}{v} & 5.56&  278.731 &    39.8 &    39.8 &     3.2 \\
 & &  281.438 &    46.9 &    46.9 &     3.2 \\
\ionx{Mg}{v} & 5.60&  276.625 &    49.9 &    46.6 &     0.5 \\
\ionx{O}{vi} & 5.64&  183.951 &    19.6 &    22.8 &     3.3 \\
 & &  184.141 &    35.8 &    39.1 &     3.3 \\
\ionx{Si}{vi} & 5.65&  246.036 &    39.0 &    39.0 &     2.4 \\
 & &  249.163 &    46.9 &    46.9 &     1.2 \\
\ionx{Mg}{vi} & 5.68&  269.020 &    32.3 &    34.5 &     1.1 \\
 & &  270.426 &    39.9 &    37.7 &     1.1 \\
\ionx{Cr}{vii} & 5.69&  202.857 &    42.9 &    42.9 &     1.5 \\
\ionx{Fe}{viii} & 5.70&  185.232 &    30.7 &    30.7 &     0.0 \\
 & &  186.627 &    45.0 &    45.0 &     1.6 \\
\ionx{Si}{vii} & 5.74&  272.685 &    41.8 &    50.6 &     1.1 \\
 & &  274.218 &    41.5 &    47.0 &     2.2 \\
 & &  275.394 &    35.9 &    44.7 &     1.1 \\
 & &  275.715 &    43.5 &    52.2 &     0.8 \\
 & &  278.484 &    37.7 &    44.2 &     2.2 \\
\ionx{Mn}{viii} & 5.74&  185.467 &    19.4 &    19.4 &     3.2 \\
\ionx{Al}{vii} & 5.76&  261.242 &    39.0 &    26.4 &     4.6 \\
 & &  259.226 &    34.7 &    22.0 &     2.3 \\
\ionx{Mg}{vii} & 5.76&  276.174 &    39.1 &    22.8 &     0.5 \\
 & &  278.426 &    35.5 &    25.8 &     1.1 \\
 & &  280.759 &    39.5 &    23.5 &     1.1 \\
\ionx{Cr}{viii} & 5.88&  205.053 &    62.9 &    62.9 &     1.5 \\
\ionx{Si}{viii} & 5.89&  250.514 &    58.7 &    76.6 &     3.6 \\
 & &  250.823 &    19.1 &    39.4 &     6.0 \\
 & &  276.873 &    24.9 &    37.9 &     2.2 \\
 & &  277.074 &    18.4 &    20.6 &     2.2 \\
\ionx{Al}{viii} & 5.92&  250.155 &    19.2 &    19.2 &     2.4 \\

\tableline
\end{tabular}
\end{center}
\end{table}

\end{document}